\documentclass[aip,jcp,graphicx,preprint]{revtex4-1}

\usepackage{graphicx}
\usepackage{dcolumn}
\usepackage{bm}
\usepackage[mathlines]{lineno}

\usepackage{amsmath,amssymb,multirow,booktabs}
\usepackage{braket,physics}
\usepackage{algorithm,algpseudocode}

\usepackage[utf8]{inputenc}
\usepackage[T1]{fontenc}
\usepackage{mathptmx}
\usepackage{etoolbox}

\makeatletter
\def\@email#1#2{%
 \endgroup
 \patchcmd{\titleblock@produce}
  {\frontmatter@RRAPformat}
  {\frontmatter@RRAPformat{\produce@RRAP{*#1\href{mailto:#2}{#2}}}\frontmatter@RRAPformat}
  {}{}
}%
\makeatother

\begin{document}

\preprint{}

\title{A Large-scale Parallel Implementation of Quasi-Four-Component Relativistic Density Functional Theory with Numeric Atom-centered Orbitals} 



\author{Wentao Zhang}
\affiliation{Thomas Lord Department of Mechanical Engineering and Materials Science, Duke University, Durham, North Carolina 27708, United States}

\author{Rundong Zhao}
\affiliation{Thomas Lord Department of Mechanical Engineering and Materials Science, Duke University, Durham, North Carolina 27708, United States}

\author{Volker Blum}
\affiliation{Thomas Lord Department of Mechanical Engineering and Materials Science, Duke University, Durham, North Carolina 27708, United States}
\affiliation{Department of Chemistry, Duke University, Durham, North Carolina 27708, United States}


\date{\today}

\begin{abstract}
We present a large-scale parallel implementation of fully relativistic density functional theory (DFT) for both molecules and periodic solids, using the quasi-four-component (Q4C) method and numeric atom-centered orbital basis sets. Our approach employs a domain decomposition method on nonuniform real-space integration grids, which enables order-N integration of the Q4C Hamiltonian matrix elements using efficient, distributed-memory and compute-parallel real-space operations. Next, we build the Hamiltonian and overlap matrices in a two-dimensional block-cyclic distribution layout. The resulting generalized eigenvalue problems are solved with the massively parallel ELPA eigenvalue solver library. We benchmark memory usage, parallel efficiency, and scalability across multiple MPI tasks and compute nodes. This algorithm extends the reach of fully relativistic DFT simulations for periodic solids, tested up to 3,383 atoms per unit cell (216,628 basis functions) and likely still well below the true reach of the implementation. As a demonstration, we calculate the fully relativistic band structure for a 3,383 atom-per-unit-cell doped hybrid organic-inorganic perovskite, (PEA)$_2$(Pb$_{1-x}$Bi$_x$)I$_4$ (PEA=phenethylammonium), showing nearly ideal scalability between 336 and 672 physical CPU cores.
\end{abstract}


\maketitle 

\section{Introduction}
Relativistic effects strongly influence materials’ physical and chemical properties, especially for the heavy-element-containing compounds. These effects are incorporated by the Dirac equation, \cite{Dirac_1928} which integrates special relativity into the framework of quantum mechanics. In line with past experience in the broader community, our group has previously published two detailed benchmark studies\cite{PhysRevMaterials.1.033803, PhysRevB.103.245144} that provide precise benchmarks of the accuracy of different relativistic methods for a broad range of inorganic compounds. For light-element based systems, the non-relativistic Schrödinger equation or the scalar relativistic (SR) \cite{Koelling_1977} variants thereof are adequate to describe many of their experimentally relevant properties. For the intermediate elements (roughly, nuclear charge $Z\ge$30), the electronic band structures and description of excited states require the inclusion of spin-orbit coupling (SOC) for physically correct results. Towards the heavy elements, there is an increasing need for self-consistent SOC and associated basis functions.  Separately, SOC and relativistic effects are essential for accurately describing electronic core levels and their excitations,\cite{Keller_2020} since the strongest relativistic effects arise close to the nucleus and govern the core-level transitions fundamental to X-ray spectroscopy. Furthermore, the magnetic coupling between atomic nuclei and the surrounding electron gas—central to Nuclear Magnetic Resonance (NMR) in heavy-element systems—relies heavily on these relativistic principles\cite{10.1098/rsta.2012.0489}. 

Over the past few decades, various electronic structure software packages implementing four-component and two-component relativistic density functional theory (DFT) have been developed for non-periodic molecules and periodic solids.  Selected examples of currently available codes and methods include the Beijing density functional (BDF)\cite{doi:10.1142/S0219633603000471,10.1063/1.5143173} code with the four-component relativistic DFT, the full-potential linear muffin-tin orbital (FP-LMTO) approach of the dirac-fp code\cite{rehn_diracs_2020}, the relativistic linearized augmented plane wave (LAPW) method in WIEN2k\cite{10.1063/1.5143061}, the four-component module of DIRAC\cite{https://doi.org/10.1002/jcc.10066}, the four-component Dirac-Kohn-Sham (4C-DKS) scheme with BERTHA\cite{10.1063/5.0002831}, the Gaussian-type orbitals based fully relativistic 4C-DKS method in RESPECT code \cite{PhysRevB.99.205103}, and the relativistic time-dependent(TD) DFT ChronusQ Package\cite{Li2019_e1436}.

One of the major bottlenecks is the lack of implementation of scalable 4C relativistic DFT calculations for large-scale molecules and solid systems. While 4C simulations are applied to small systems, they remain much less available for large, complex systems in production calculations. Furthermore, much of the functionality implemented in scalar relativistic (SR) codes \cite{VISSCHER1999357, 10.1063/1.1329891} is harder to access in 4C variants. Consequently, most production calculations currently avoid direct 4C treatments.

In this paper, we develop a distributed-memory parallel 4C approach using the quasi-four-component (Q4C) density functional theory and numeric atom-centered orbital (NAO) basis sets. In essence, the overall method builds upon the seminal, overlapping atom-centered grid based integration scheme initially pioneered by Axel Becke\cite{10.1063/1.454033} and extends its reach to system sizes that are, to our knowledge, unprecedented for four-component relativistic DFT. We implemented the algorithms in the FHI-aims \cite{BLUM20092175, HAVU20098367,abbott2026roadmap}, an all-electron electronic structure theory code, building on a previous Q4C implementation by some of us\cite{PhysRevB.103.245144} that had been proven up to 118 atoms in a complex material\cite{abbott2026roadmap} but that is limited towards larger system sizes due to growing per-processor memory usage. In the present work, this restriction is lifted and much larger system sizes are achieved using a modified domain-decomposition method on real-space integration grids, with order-N scaling for the integration of the Hamiltonian and overlap matrices and for the grid-based electron density update. This development makes it possible to perform large-scale fully relativistic DFT calculations on both non-periodic and periodic systems, leveraging the efficient parallelism of central processing unit (CPU) based HPC clusters with low latency interconnect. The maximum system size demonstrated is a 3,383 atom, lead (Pb) based layered perovskite supercell doped with the neighboring element Bi, reflecting the realistic needs of a past study of the physics of such systems in our group.\cite{PRXEnergy.2.023010} This demonstration is carried out with 672 physical CPU cores, indicative of the HPC hardware available for this study. However, with larger hardware resources, attainable system sizes and scalability may actually exceed the concrete test cases demonstrated in this work.

This paper is organized as follows. In Section 2, we introduce the theoretical formulation of the four-component Dirac-Kohn-Sham theory and the Q4C approach with numerical atom-centered orbitals for both non-periodic and periodic systems. Section 3 describes the implementation details of the large-scale parallel Q4C algorithm. In section 4, we present the timing and scaling benchmarking results for this algorithm, as well as the electronic band structures with different levels of relativistic treatment for large hybrid organic-inorganic perovskites. Finally, section 5 provides conclusions and future perspectives.

\section{Theory}
\subsection{Four-component Dirac-Kohn-Sham theory}
The Dirac equation provides the foundation for incorporating relativistic effects into the electronic-structure theory. Relativistic density functional theory (RDFT) \cite{PhysRevB.7.1912, Rajagopal_1978} extends this framework to the Kohn–Sham formalism, resulting in the Dirac–Kohn–Sham (DKS) equation \cite{dyall2007relativistic} for the single-particle four-component wave functions $\phi_i$, which is the starting point of this work:
\begin{equation}
\left(
\begin{array}{cc}
\begin{pmatrix}
V_{\text{eff}} & 0\\
0 & V_{\text{eff}}
\end{pmatrix}
&
c\,\boldsymbol{\sigma}\cdot\hat{\boldsymbol{p}}
\\[6pt]
c\,\boldsymbol{\sigma}\cdot\hat{\boldsymbol{p}}
&
\begin{pmatrix}
V_{\text{eff}}-2mc^2 & 0\\
0 & V_{\text{eff}}-2mc^2
\end{pmatrix}
\end{array}
\right)
\begin{pmatrix}
\phi_{1,i}\\
\phi_{2,i}\\
\phi_{3,i}\\
\phi_{4,i}
\end{pmatrix}
=
\epsilon_i\,
\begin{pmatrix}
\phi_{1,i}\\
\phi_{2,i}\\
\phi_{3,i}\\
\phi_{4,i}
\end{pmatrix} \, .
\label{DKS_equation}
\end{equation}
The effective potential $V_{\text{eff}}$ consists of the exchange-correlation potential $V_{\text{xc}}$, the Hartree potential $V_{\text{H}}$, and the electron-nuclear attraction $V_{\text{N}}$. $\phi_{j,i}$ ($j$=1,$\dots$,4) are complex-valued, scalar functions. The quantities $\epsilon_i$ are the eigenvalues, $\hat{\boldsymbol{p}}$ is the momentum operator, and $\boldsymbol{\sigma}$ is the vector of three Pauli spin matrices. In Eq.(\ref{DKS_equation}), the Dirac-Coulomb Hamiltonian \cite{dyall2007relativistic} is utilized without inclusion of the orbital current terms \cite{rehn_diracs_2020}, and the relativistic corrections to the Coulomb interactions have been neglected (see, e.g., Ref. \cite{autschbach_2012} for a compact review). The wave functions are four-component (4C) spinors, which can be decomposed into a first two-component spinor $\phi^L_{i}$, termed the "large component", and a second two-component spinor $\phi^S_{i}$, referred to as the "small component", which can be summarized as
 \begin{equation}
\phi_{i}=\begin{pmatrix}
\phi_{1,i}\\
\phi_{2,i}\\
\phi_{3,i}\\
\phi_{4,i}
\end{pmatrix} =\begin{pmatrix}
\phi^L_{i}\\
\phi^S_{i}\\
\end{pmatrix} \, .
\label{spinors}
 \end{equation}
Under the no-pair approximation\cite{10.1063/1.4959452}, the Dirac-Coulomb Hamiltonian is projected to eliminate the negative energy states corresponding to positrons. Thereby, the total energy of the many-electron system can be evaluated as
\begin{align}
E_{\text{tot}}  &= \sum_{i=1}^{N_{\text{states}}}f_i\epsilon_i-\int d^3r[n(\mathbf{r})V_{\text{xc}}(\mathbf{r})]+E_{\text{xc}}[n] -\frac{1}{2}\int d^3r[n(\mathbf{r})V_{\text{H}}(\mathbf{r})]+E_{\text{nuc-nuc}}
\end{align}
where $f_i$ denotes the occupation numbers of the orbitals, $E_{\text{nuc-nuc}}$ stands for the internuclear repulsion, and $E_{\text{xc}}$ is the exchange-correlation functional. For $E_\text{xc}$ in particular, we here adopt standard scalar-relativistic density functional approximations. The electron density $n(\mathbf{r})$ is evaluated as the superposition of the large and small component densities:
\begin{equation}
    n(\mathbf{r}) = n^{L}(\mathbf{r}) + n^{S}(\mathbf{r}) = \sum_if_i\abs{\phi^L_i(\mathbf{r})}^2 + \sum_if_i\abs{\phi^S_i(\mathbf{r})}^2 \, .
\end{equation}


\subsection{Q4C method with numerical atom-centered basis functions}
From the DKS equation, the exact relation between the large and small components of the wave functions is
\begin{equation}
    \phi_i^S(\mathbf{r}) = \frac{c}{2mc^2+\epsilon_i-V_{\text{eff}}(\mathbf{r})}(\boldsymbol{\sigma}\cdot\hat{\boldsymbol{p}})\phi_i^L(\mathbf{r}) \, .
\label{Exact relation}
\end{equation}
The small component wave functions $\phi_i^S$ exhibit significant magnitude only in the close vicinity of the nucleus \cite{PhysRevB.103.245144}, where the effective potential $V_{\text{eff}}$ is on the order of $2mc^2$, rendering the relatively minor perturbations imposed by the chemical environment negligible. Consequently, the ratio between the large and small components near the nucleus can be fixed to its precomputed value in the corresponding spherical free atom with high accuracy \cite{10.1063/1.3159445, 10.1063/1.2137315, 10.1063/1.473860}. The Quasi-four-component (Q4C) approach \cite{10.1063/1.2222365, 10.1063/1.3159445, PhysRevB.103.245144} exploits this idea by expanding the molecular wave functions using free-atom-like basis functions. In this formalism, the large and small component basis functions share the same set of coefficients $C^+$, allowing the molecular 4C wavefunctions to be expressed as a linear combination of atom-centered 4C spinor basis functions:
\begin{equation}
    \phi_i = \begin{pmatrix}
        \phi^L_i \\
        \phi^S_i \\
    \end{pmatrix}
    =\sum_{\mu}^{N_{\text{spinor}}}(C^+)_{\mu i}
    \begin{pmatrix}
        \chi^L_\mu \\
        \chi^S_\mu \\
    \end{pmatrix} .
\label{LC4CS}
\end{equation}
With the positive electronic energy states only, the number of 4C spinor basis functions is $N_{\text{spinor}}$. Inserting Eq. (\ref{LC4CS}) into the exact relation of Eq. (\ref{Exact relation}) for a state $i$, one can derive an ideal relation for each of the large and small components' spinor basis functions
\begin{equation}
\chi^S_\mu = \hat{K}\chi^L_\mu = \frac{c}{2mc^2+\epsilon_{i}-V_{\text{eff}}(\mathbf{r})}(\boldsymbol{\sigma}\cdot\hat{\boldsymbol{p}})\chi^L_\mu
\label{exactK}
\end{equation}
However, in order to derive all states from a single (not itself state-dependent) Hamiltonian matrix, the basis functions used to express the small component cannot explicitly depend on a particular state $i$. In our approach, the definition of operator $\hat{K}$ in Eq. (\ref{exactK}) therefore depends on the types of basis functions associated with an atomic center $A$, the corresponding free-atom orbital energy $\epsilon_{A,\mu}$, and free-atom radial potential $V_{A}$. 
\begin{subequations}
\begin{align}
\hat{K} &= \frac{c}{2mc^2+\epsilon_{A,\mu}-V_{\text{A}}(\mathbf{r})}\boldsymbol{\sigma}\cdot\hat{\boldsymbol{p}} \, ,\\
\hat{K} &= \frac{c}{2mc^2-V_{\text{A}}(\mathbf{r})}\boldsymbol{\sigma}\cdot\hat{\boldsymbol{p}} \, .
\end{align}
\end{subequations}
For basis functions derived from the free-atom eigenproblem, $\hat{K}$ adopts the exact form of Eq (8a), which is a condition known as the atomic balance \cite{https://doi.org/10.1002/qua.560400816}. The rationale for this choice is that an atomic core radial function (for which $\epsilon_{A,\mu}$ deviates appreciably from zero, compared to $2mc^2$) in a multi-atom system is expected to be very similar to its free-atom counterpart. Conversely, that free-atom counterpart will therefore not contribute much to other states. For all other basis functions, we replace $\epsilon_i$ with zero in Eq. (\ref{exactK}) and use the form of Eq (8b), similar to the atomic ZORA approximation as defined in Ref. \cite{BLUM20092175}. A more detailed discussion can be found in Section II B of Ref. \citenum{PhysRevB.103.245144}.

The relativistic 4C spinor basis functions can be written in a separable form as a product of radial and spin-angular parts
\begin{equation}
    \chi(\mathbf{r}) =\begin{pmatrix}
        \chi^L(\mathbf{r}) \\
        \chi^S(\mathbf{r}) \\
    \end{pmatrix} = \frac{1}{r}\begin{pmatrix}
        P(r)\Upsilon_{\kappa,m_j}(\Omega) \\
        iQ(r)\Upsilon_{-\kappa,m_j}(\Omega) \\
    \end{pmatrix} \, .
\end{equation}
$P(r)$ and $Q(r)$ are radial functions of large and small components of the basis function, respectively, and the imaginary number $i$ is introduced to keep the radial function $Q(r)$ real-valued. The $\Omega=(\theta,\phi)$ denotes the angles in spherical coordinates relative to the atomic center of the basis function. $\kappa$ (see below) and $m_j$ are the usual quantum numbers of the Dirac equation for a spherically symmetric potential.  The spin-angular function $\Upsilon_{\kappa,m}$ is represented as the coupled functions between spin functions using the Clebsch-Gordan coefficients $C_\alpha, C_\beta$ and the complex spherical harmonics, which gives
\begin{equation}
  \Upsilon_{\kappa,m_j}= 
    \begin{pmatrix}
      C_\alpha \Tilde{Y}_{l}^{m_j-\tfrac12}(\Omega)\\
      C_\beta \Tilde{Y}_{l}^{m_j+\tfrac12}(\Omega)
  \end{pmatrix}=
  \frac{1}{\sqrt{2l+1}}
    \begin{pmatrix}
    a\sqrt{l + am_j + \tfrac12}\;\Tilde{Y}_{l}^{m_j-\tfrac12}(\Omega)\\
    \sqrt{l - am_j + \tfrac12}\;\Tilde{Y}_{l}^{m_j+\tfrac12}(\Omega)
  \end{pmatrix} \, .
\end{equation}
This indicates that each of the large and small components of the spinor basis functions has a different spin-angular function associated with its radial part. The detailed formulas between Clebsch-Gordan coefficients and complex spherical harmonics are summarized in appendix \ref{CG}.  $\kappa$ is the Dirac quantum number, which is determined by the total angular momentum $j$ and orbital angular momentum $l$.
\begin{equation}
    \kappa=-a(j+\frac{1}{2})=\begin{cases}
-l-1,  & j=l+\frac{1}{2} \, ,\\
l,   & j=l-\frac{1}{2} \, .
\end{cases}
\end{equation}
Here, $a=2(j-l)$ is a sign factor with values $\pm1$, $a$ is $-$1 when $\kappa>0$ and $+$1 when $\kappa<0$. A detailed explanation of the quantum numbers can be found in Ref. \cite{dyall2007relativistic}. 

One particular choice for representing spinor basis functions in a scalar basis expansion is to employ numeric atom-centered orbital (NAO) basis functions. In our work, the relativistic large and small components of scalar basis functions then take 
the NAO form
\begin{subequations}
\begin{align}
    \varphi^L_{t}(\mathbf{r})&=\frac{P(r)}{r}Y_{l,m}(\Omega) \, ,
\\
    \varphi^S_{t}(\mathbf{r})&=\frac{iQ(r)}{r}Y_{\tilde{l},\tilde{m}}(\Omega) \, ,
\end{align} \label{scalarbasis}
\end{subequations}
where the radial functions $P(r)$ and $Q(r)$ are derived by numerically solving the radial Dirac equation for the free atom with the DFTATOM solver \cite{CERTIK20131777} on one-dimensional logarithmic grids. The $Y_{l,m}$, $Y_{\tilde{l},\tilde{m}}$ denote the real-valued spherical harmonics for large and small components scalar basis functions. The Condon-Shortley phase convention is used for the real spherical harmonics; refer to appendix \ref{RSH} for explicit definitions. The complex spherical harmonics $\Tilde{Y}_{l}^{m}$  can be expressed in terms of the real spherical harmonics $Y_{l,m}$ via the inverse transformation:
\begin{equation}
\Tilde{Y}_{l}^{m}(\Omega) = \begin{cases}
\frac{1}{\sqrt{2}} \left( Y_{l,|m|}(\Omega) - iY_{l,-|m|}(\Omega) \right) & \text{if } m < 0, \\
Y_{l,0}(\Omega) & \text{if } m = 0, \\
\frac{(-1)^{m}}{\sqrt{2}} \left( Y_{l,|m|}(\Omega) + iY_{l,-|m|}(\Omega) \right) & \text{if } m > 0.
\end{cases}
\end{equation}

In practice, we never explicitly build these complex spherical harmonics in the spin-angular functions. Instead, by folding this fixed unitary transformation directly into the Clebsch-Gordan coupling, the two-component large and small spinor basis functions are evaluated during the numerical integration as linear combinations of the scalar basis functions,
\begin{subequations}
\begin{align}
    \chi^L_{\mu}(\mathbf{r})&=\sum^{N^L_\text{scalar}}_{t} G^L_{t\mu}\varphi^L_t(\mathbf{r})=\sum^{N^L_\text{scalar}}_{t}\begin{pmatrix}
         C^L_\alpha\varphi^L_{t,\alpha}(\mathbf{r})\\
        C^L_\beta\varphi^L_{t,\beta}(\mathbf{r})
\end{pmatrix} \, ,
\\
    \chi^S_{\mu}(\mathbf{r})&=\sum^{N^S_\text{scalar}}_{t} G^S_{t\mu}\varphi^S_t(\mathbf{r})=\sum^{N^S_\text{scalar}}_{t}\begin{pmatrix}
         C^S_\alpha\varphi^S_{t,\alpha}(\mathbf{r})\\
        C^S_\beta\varphi^S_{t,\beta}(\mathbf{r})
\end{pmatrix} \, .
\end{align}
\end{subequations}
The number of large and small component scalar basis functions is represented as $N^L_\text{scalar}$ and $N^S_\text{scalar}$, with $N^L_\text{scalar} < N^S_\text{scalar}$, since the small component spin-angular functions carry $\tilde{l} = l \pm 1$. The transformation matrices $G^L_{t\mu}$ and $G^S_{t\mu}$ contain the Clebsch--Gordan coefficients that couple the scalar basis functions to the spinor representation, with dimensions $N^L_\text{scalar} \times N_\text{spinor}$ for the large component and $N^S_\text{scalar} \times N_\text{spinor}$ for the small component.

\subsection{Integration and density evaluation}
It should be emphasized that while the formulas below are presented using the two-component spinor basis functions for compact notation, the implementation evaluates these quantities using the real-valued scalar NAO basis functions. Substituting Eq. (6) into Eq. (1) yields the generalized eigenvalue problem for electronic states only
\begin{equation}
    H^+C^+=S^+C^+E^+ \, .
\end{equation} 
The corresponding Hamiltonian and overlap matrix elements in the spinor basis functions form are
\begin{subequations}
\begin{align}
H_{\mu\nu}^+
&=
\Braket{
\begin{matrix}\chi_\mu^{L}\\ \chi_\mu^{S}\end{matrix}
\Bigg|
\begin{matrix}
V_{\text{eff}} & c\,\boldsymbol{\sigma}\!\cdot\!\mathbf{p} \\
c\,\boldsymbol{\sigma}\!\cdot\!\mathbf{p} & V_{\text{eff}}-2mc^{2}
\end{matrix}
\Bigg|
\begin{matrix}\chi_\nu^{L}\\ \chi_\nu^{S}\end{matrix}
} \notag \\
&= \Braket{\chi_{\mu}^{L}|c\,\boldsymbol{\sigma}\!\cdot\!\mathbf{p}|\chi_{\nu}^{S}}
+ \Braket{\chi_{\mu}^{S}|c\,\boldsymbol{\sigma}\!\cdot\!\mathbf{p}|\chi_{\nu}^{L}} \notag \\
&
+ \Braket{\chi_{\mu}^{L}|V_{\text{eff}}|\chi_{\nu}^{L}}
+ \Braket{\chi_{\mu}^{S}|V_{\text{eff}}-2mc^{2}|\chi_{\nu}^{S}} \, ,
\\
S_{\mu\nu}^+
&= \Braket{\chi_\mu^{L}|\chi_\nu^{L}}
+ \Braket{\chi_\mu^{S}|\chi_\nu^{S}} \, .
\end{align}
\end{subequations}

Direct evaluation of this exact form of the Hamiltonian matrix elements on integration grids would lead to very slow numerical convergence with integration grid density, due to the Coulomb singularity and the singularity of Dirac radial functions near the nucleus \cite{dyall2007relativistic, PhysRevB.103.245144, CERTIK20131777}. To avoid these numerical complications, we exploit the construction of numeric atom-centered orbital basis functions. Because these basis functions are generated as exact numerical solutions to the atomic Dirac equation on dense, one-dimensional logarithmic grids, the action of the kinetic energy operator can be recast analytically as
\begin{subequations}
\begin{align}
c\,\boldsymbol{\sigma}\!\cdot\!\mathbf{p}\,\ket{\chi^{S}_{\nu}} &= \bigl(\varepsilon_{\nu}-V_{\nu}\bigr)\,\ket{\chi^{L}_{\nu}} \, ,\\
c\,\boldsymbol{\sigma}\!\cdot\!\mathbf{p}\,\ket{\chi^{L}_{\nu}} &= \bigl(\varepsilon_{\nu}-V_{\nu}+2mc^{2}\bigr)\,\ket{\chi^{S}_{\nu}} \, .
\end{align}
\end{subequations}
For each type of orbital, we apply the corresponding eigenvalue $\varepsilon_{\nu}$ and its defining potential $V_{\nu}(r)$ (e.g., free-atom-like, hydrogen-like, or free-ion-like)\cite{BLUM20092175}:
\begin{equation}
V_{\nu}(r) = 
\begin{cases} 
V_{\text{atom}}(r)  & \text{for free-atom like orbitals,} \\ 
V_{\text{hydro}}(r), & \text{for hydrogen-like orbitals,} \\ 
V_{\text{ion}}(r), & \text{for ionic orbitals.} 
\end{cases}
\end{equation}

Substituting these relations allows us to combine the kinetic and potential energy contributions. Since the effective potential $V_{\text{eff}}$ and the atomic potential $V_\text{atom}$ exhibit identical singular behavior near the nucleus, the steep contributions especially of the important free-atom-like basis functions cancel out. This results in a numerically smooth function that can be integrated with high precision.
\begin{equation}
    H_{\mu\nu}^+=
\Braket{\chi_\mu^{L}|\bigl(\varepsilon_{\nu}-V_{\nu}+V_{\text{eff}}\bigr)\,|\chi^{L}_{\nu}}
+\Braket{\chi_\mu^{S}|\bigl(\varepsilon_{\nu}-V_{\nu}+V_{\text{eff}}\bigr)\,|\chi^{S}_{\nu}} \, .
\end{equation}

After constructing the Hamiltonian and overlap matrices, we then solve the generalized eigenvalue problem with the ELPA eigensolver\cite{marek_elpa_2014} via the ELSI infrastructure\cite{YU2018267, YU2020107459}. The electron density from a given set of Kohn–Sham eigenvectors $C^{+}_{\mu l}$ can be derived with a density matrix-based update method.\cite{HAVU20098367} The density matrix in the spinor basis form is:
\begin{equation}
    D_{\mu\nu}=\sum_l^{N_{\text{occ}}} f_l{C^{+}_{\mu l}}^{*}C^{+}_{\nu l} \, .
\end{equation}
 $N_{\text{occ}}$ is the number of occupied Dirac-Kohn–Sham eigenstates and $f_l$ is the occupation number of eigenstate $l$. Using the resulting spinor density matrix $D_{\mu\nu}$,  the electron density $n(\mathbf{r})$ is evaluated as
\begin{equation}
\begin{aligned}
n(\mathbf r)
&= \operatorname{Re}\biggl[\,\sum_{\mu,\nu}^{N_{\mathrm{spinor}}} D_{\mu\nu} ({\chi^{L}_{\mu}}^{*}\chi^{L}_{\nu}+{\chi^{S}_{\mu}}^{*}\chi^{S}_{\nu})\biggr] \, .
\end{aligned}
\end{equation}
Note that this density expression is formulated in terms of the spinor basis functions, whereas the actual implementation in terms of scalar NAO functions is detailed in Section IIIA.

\subsection{Extension to periodic systems}
For periodic systems, the $\mathbf{k}$-space-dependent Bloch-like generalized spinor basis functions are defined from the localized real-space spinor basis functions that are centered in the unit cells labeled $M$ with translation vectors $\mathbf{T}_M$, which gives
\begin{equation}
    \psi_{\mu,\mathbf{k}}(\mathbf{r}) =
\sum_M \exp[{{i\mathbf{k}\cdot \mathbf{T}_M}}]
    \begin{pmatrix}
        \chi^L_\mu(\mathbf{r}-\mathbf{T}_M) \\
        \chi^S_\mu(\mathbf{r}-\mathbf{T}_M) \\
    \end{pmatrix} \, . \label{Bloch}
\end{equation}
Here, $\mathbf k$ is the crystal momentum and the sum over $M$ runs over all periodic images of the reference unit cell, which extend infinitely across the crystal. In the definition of Eq. (\ref{Bloch}), the set of basis functions $\mu$ only encompasses those basis functions whose atomic centers are associated with the zeroth (0-th) unit cell of the crystal. However, all matrix elements can be cast in a form normalized to the volume of a single reference unit cell, as shown in detail in Eqs.~(27)–(31) of Ref.~\citenum{KNUTH201533} and Eqs.~(11)–(12) of Ref.~\citenum{HUHN2020107314}. For a general operator $\hat {O} $, the portion of the integral that lies within the volume of the 0-th unit-cell (uc) is
\begin{equation}
O^{uc}_{\mu\nu}
=\Braket{\chi^{X}_{\mu}\big|\hat O\big|\chi^{X}_{\nu}}_{uc}
=\int_{uc} \chi^{X}_{\mu}(\mathbf r)\,\hat O\,\chi^{X}_{\nu}(\mathbf r)\, d^{3}\mathbf r,
\qquad \text{for } X = L, S \, . \label{uc-int}
\end{equation}
In the periodic case, the indices $\mu$ and $\nu$ range from $(1,...,N^{\mathrm{uc}}_{\mathrm{spinor}})$. Crucially, and unlike for Eq. (\ref{Bloch}), in Eq. (\ref{uc-int}) these indices do not only run over the localized real-space basis functions whose atomic centers are associated with the 0-th unit cell. Rather, the relevant set of $N^{\mathrm{uc}}_{\mathrm{spinor}}$ basis functions corresponds to those spinor basis functions that have a non-zero piece within the central (0-th) unit cell, regardless of whether their atomic center is associated with the 0-th or with another unit cell. The partial integrals $O^{uc}_{\mu\nu}$ are, however, only evaluated within the volume of the 0-th unit cell, while the Bloch-like nature of the extended basis functions $\psi_{\mu,\mathbf{k}}(\mathbf{r})$ in Eq. (\ref{Bloch}) is separately recovered through the inclusion of phase factors $\exp[{i\mathbf{k}\cdot(\mathbf{T}_{M(\nu)}-\mathbf{T}_{M(\mu)}}]$. $M(\mu)$ and $M(\nu)$ denote the unit cell where the spinor basis functions $\mu$ and $\nu$ are centered. The resulting $\mathbf{k}$-dependent matrix elements in the representation of extended Bloch-like spinor basis $\psi_{\mu,\mathbf{k}}$ can then be obtained as:
\begin{equation}
O_{\mu\nu}(\mathbf{k})={\sum_{\mu^\prime,\nu^\prime}}^{'} \exp[{i\mathbf{k}\cdot (\mathbf{T}_{M(\nu^\prime)}-\mathbf{T}_{M(\mu^\prime})}]O^{uc}_{\mu^\prime\nu^\prime} \, .
\label{restrictedsum}
\end{equation}
The restricted double sum ${\sum_{\mu^\prime,\nu\prime}}^{'}$ runs over all real-space spinor basis functions $\mu^\prime$ and $\nu^\prime$ that have a non-zero support in the 0-th unit cell and where basis function $\mu^\prime$ is a periodic image of basis function $\mu$ in the 0-th unit cell and $\nu^\prime$ is a periodic image of basis function $\nu$ in the 0-th unit cell, respectively. To be specific and using the same convention for the indices, the large and small components of the Hamiltonian matrix elements in periodic boundary conditions are defined by
\begin{subequations}
\begin{align}
H^{L}_{\mu\nu}(\mathbf{k}) &=
  {\sum_{\mu^\prime,\nu^\prime}}^{'} \exp[{i\mathbf{k}\cdot (\mathbf{T}_{M(\nu^\prime)}-\mathbf{T}_{M(\mu^\prime)})}]\Braket{\chi_{\mu^\prime}^{L}|\bigl(\varepsilon_{\nu^\prime}-V_{\nu^\prime}+V_{\text{eff}}\bigr)\,|\chi^{L}_{\nu^\prime}}_{uc} \, ,
\\
H^{S}_{\mu\nu}(\mathbf{k}) &=
  {\sum_{\mu^\prime,\nu^\prime}}^{'} \exp[{i\mathbf{k}\cdot (\mathbf{T}_{M(\nu^\prime)}-\mathbf{T}_{M(\mu^\prime)})}]\Braket{\chi_{\mu^\prime}^{S}|\bigl(\varepsilon_{\nu^\prime}-V_{\nu^\prime}+V_{\text{eff}}\bigr)\,|\chi^{S}_{\nu^\prime}}_{uc} \, .
\end{align}
\end{subequations}
Similarly, the overlap matrix elements in periodic boundary conditions for the large and small components are given by
\begin{subequations}
\begin{align}
S^{L}_{\mu\nu}(\mathbf{k}) &=
  {\sum_{\mu^\prime,\nu^\prime}}^{'} \exp[{i\mathbf{k}\cdot (\mathbf{T}_{M(\nu^\prime)}-\mathbf{T}_{M(\mu^\prime)})}]
  \bra{\chi^L_{\mu^\prime}}\ket{\chi^L_{\nu^\prime}}_{uc}  \, ,\\
S^{S}_{\mu\nu}(\mathbf{k}) &=
  {\sum_{\mu^\prime,\nu^\prime}}^{'} \exp[{i\mathbf{k}\cdot (\mathbf{T}_{M(\nu^\prime)}-\mathbf{T}_{M(\mu^\prime)})}]
  \bra{\chi^S_{\mu^\prime}}\ket{\chi^S_{\nu^\prime}}_{uc} \, .
\end{align} 
\end{subequations}
After integrating all these matrix elements for the Hamiltonian and overlap matrices, we then solve the $\mathbf{k}$-dependent generalized eigenvalues problem, $H^{+}(\mathbf{k})C^{+}(\mathbf{k})=S^{+}(\mathbf{k})C^{+}(\mathbf{k})E^{+}(\mathbf{k})$, using the ELPA eigensolver via ELSI.

In periodic systems, the sum over occupied states to form the density matrix implicitly includes the summation over $\mathbf{k}$-points and the corresponding phase factors.
The $\mathbf{k}$-dependent coefficients and phase factors are absorbed into the spinor density matrix
\begin{equation}
    D_{\mu\nu} = \sum_{\mathbf{k}} \exp[{i\mathbf{k}\cdot(\mathbf{T}_{M'}-\mathbf{T}_M)}] \sum_l^{N_{\text{occ}}} f_l(\mathbf{k}) {C^{+}_{\mu l}(\mathbf{k})}^{*} C^{+}_{\nu l}(\mathbf{k})
\end{equation}
where $f_l(\mathbf{k})$ is the occupation number for the state $l$ at point $\mathbf{k}$. The electron density $n(\mathbf{r})$ is evaluated as
\begin{equation}
\begin{aligned}
n(\mathbf r) &= \operatorname{Re}\biggl[\,\sum_{\mu,\nu}^{N^{\mathrm{uc}}_{\mathrm{spinor}}} D_{\mu\nu}({\chi^{L}_{\mu}}^{*}\chi^{L}_{\nu}+{\chi^{S}_{\mu}}^{*}\chi^{S}_{\nu})\biggr] \, .
\end{aligned}
\end{equation}

\section{Implementation}
\subsection{Locally-indexed real-space domain decomposition}

Locally-indexed real-space domain decomposition is an efficient integration algorithm for spatially localized basis functions, which enables both memory and computational parallelization of real-space operations,\cite{HUHN2020107314} e.g., the real-space integration of the Hamiltonian and overlap matrix elements and the density matrix-based electron density update. The term ``locally-indexed'' means that, during the integration step, the integration grid is split into localized subsections that are assigned to separate MPI tasks. Each MPI task only handles (``indexes'') the matrix elements associated with the subset of non-zero basis functions on the subsection of the integration grid that is assigned to it, limiting the per-processor memory required to store the partial integrals associated with each MPI task. For scalar-relativistic calculations and post-self-consistent SOC, this algorithm has long ensured scalability to extreme system sizes within the FHI-aims code.\cite{BLUM20092175, HUHN2020107314, Kokott_2024} In this section, we extend this approach to the relativistic Q4C method with NAO basis functions. We here recapitulate the essentials of the locally-indexed real-space domain decomposition algorithm as described in previous papers \cite{BLUM20092175, HAVU20098367, KNUTH201533, HUHN2020107314} and then address the specific modifications needed for the four-component case.

\begin{figure}
    \centering
    \includegraphics[width=0.85\linewidth]{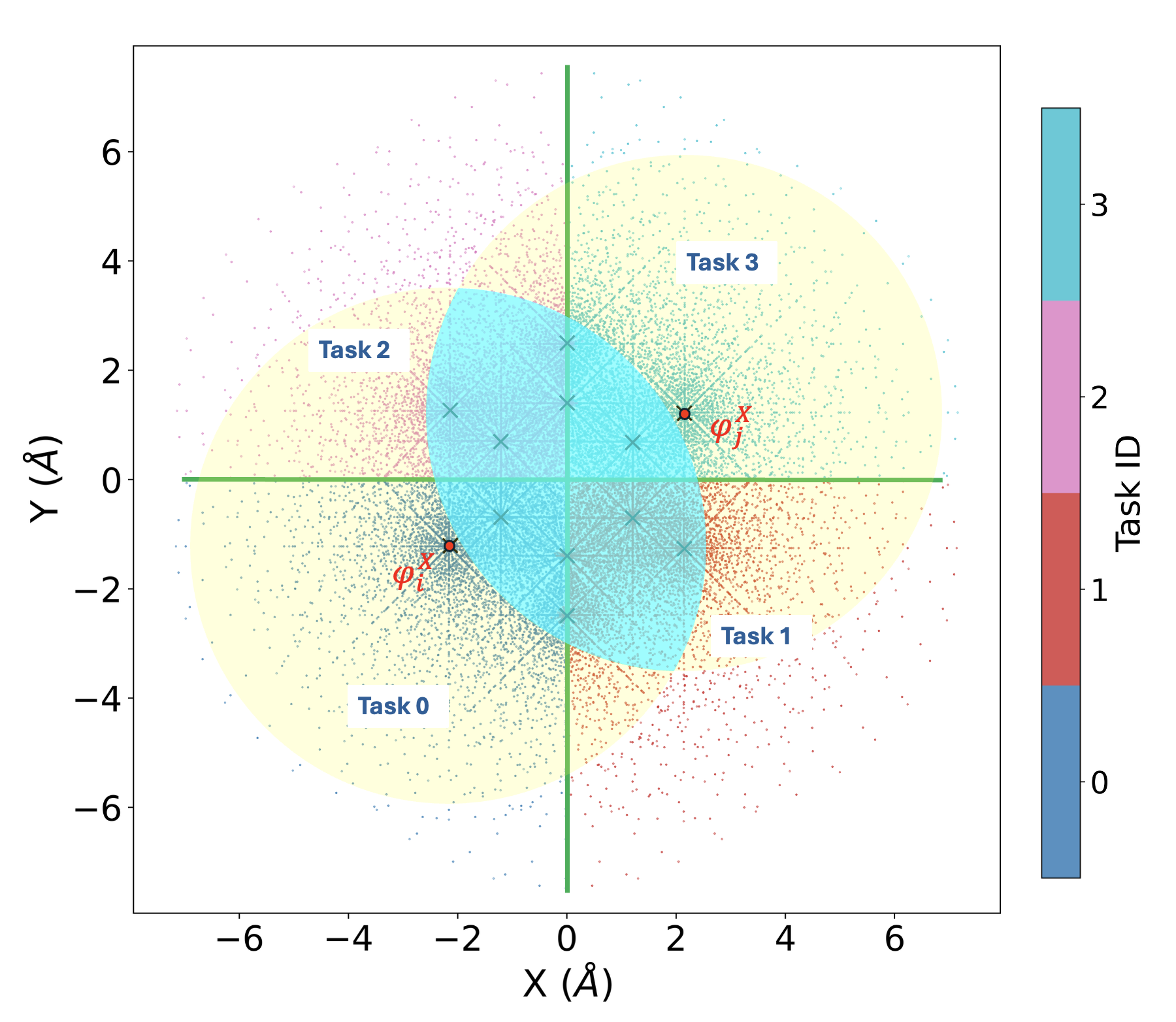}
    \caption{Real-space grid partitioning for a benzene molecule. Colored points represent the local set of integration points owned by each MPI task (Tasks 0--3). Yellow circles indicate the extent of two pairs of large and small component scalar basis functions $\varphi_i^X$ and $\varphi_j^X$, centered at two distinct atoms. The cyan region is the spatial overlap of these basis function supports, which defines the active integration volume for evaluating matrix elements. Black crosses denote atomic sites. Units are in Angstrom (\AA).} 
    \label{fig:partition_molecule}
\end{figure}

\begin{figure}
    \centering
    \includegraphics[width=1.0\linewidth]{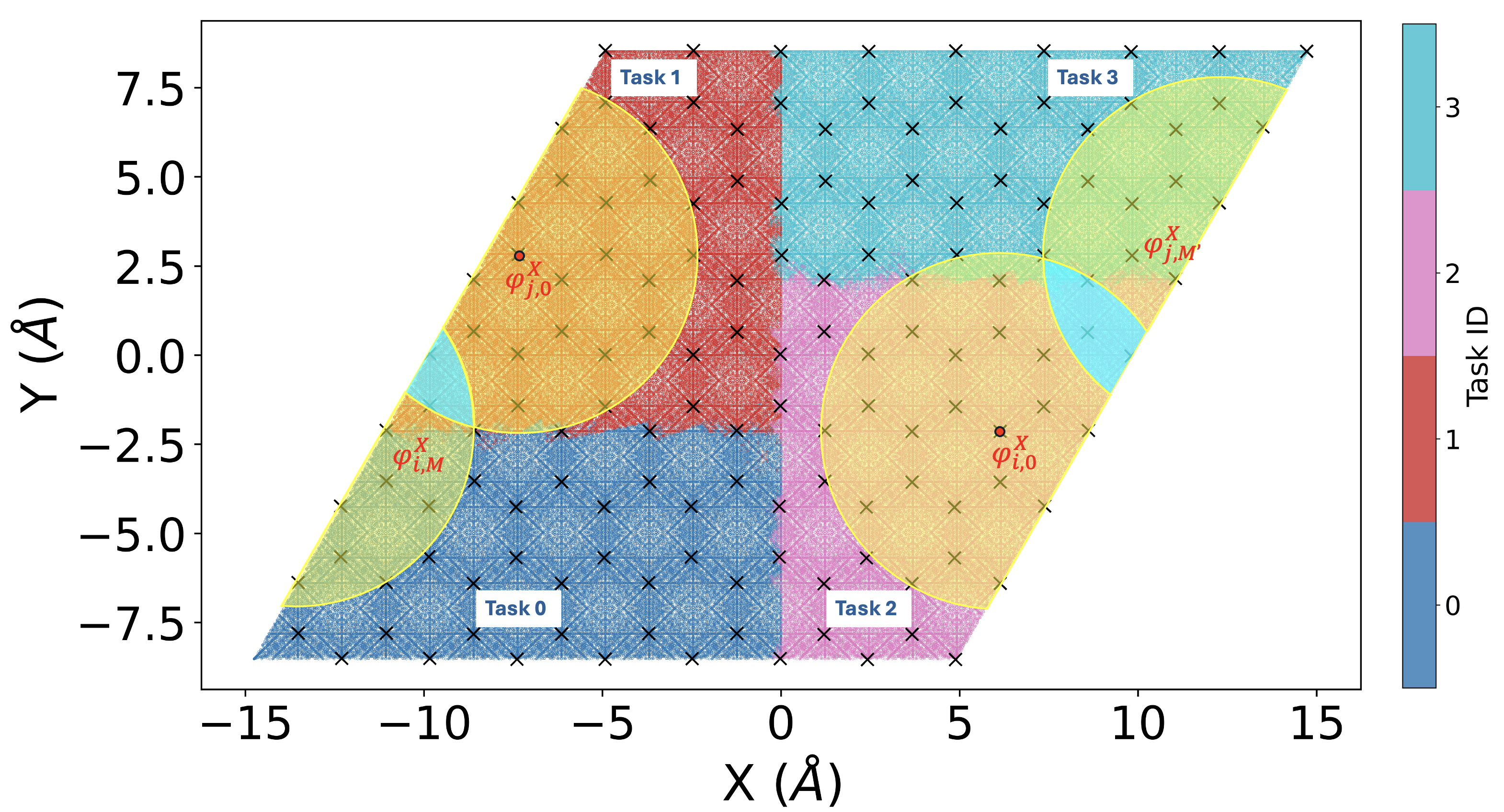}
    \caption{Real-space grid partitioning for an $8\times8$ periodic graphene supercell. The rhombic boundaries are the central (0-th) unit cell. Colored points represent the local set of integration points owned by each MPI task (Tasks 0--3). Yellow circles denote the spatial support of the localized scalar basis functions, which originate from atoms located within the 0-th cell ($\varphi_{i,0}^X$, $\varphi_{j,0}^X$) or from neighboring periodic images ($\varphi_{i,M}^X$, $\varphi_{j,M'}^X$). The cyan areas highlight the spatial overlap of these basis functions strictly within the 0-th unit cell. This defines the active integration volume that yields non-zero contributions to the unit-cell matrix elements $F^{uc}_{ij}$. Black crosses denote atomic sites. Units are in Angstrom (\AA).}
    \label{fig:partition_solids}
\end{figure}

Real-space numerical integrations are performed on a superposition of overlapping atom-centered grids.\cite{10.1063/1.454033, 10.1063/1.458452, BLUM20092175} Each grid consists of a set of spherical radial shells of grid points around each atom, with angular distributions of grid points on each shell described by Lebedev \cite{LEBEDEV197544, LEBEDEV197610}and Delley \cite{10.1063/1.458452}. For the example of the benzene molecule, these grids are shown in a projection as colored points in Fig \ref{fig:partition_molecule}. An analogous image is shown for a periodic graphene monolayer in Fig \ref{fig:partition_solids}, in which the grid points that are actually used are all located in the central (0-th) unit cell. In both images, grid points associated with different MPI tasks are shown in different colors and the non-zero domains of two exemplary, overlapping basis functions are overlaid as circles on top of the integration grid (more below). To accurately integrate matrix elements using these overlapping atom-centered grids, a partition-of-unity approach is applied \cite{10.1063/1.454033, HUHN2020107314}. Specifically, the modified Stratmann partitioning scheme \cite{STRATMANN1996213} as defined in Ref.\cite{KNUTH201533} is employed, where the integration weights $w(r)$ are calculated on the fly. 

Crucially, the actual integrations of any matrix elements are performed using the scalar-relativistic versions of the basis functions, $\varphi^{X}_i$, as defined in Eq. (\ref{scalarbasis}), not the spinor form $\chi^{X}_\mu$ (where $X$=$L$ or $S$). Using the scalar form makes it possible to reuse the complex-valued spherical harmonics and other code associated with actual scalar-relativistic integrals. In other words, scalar relativity, spin-orbit coupling and Q4C can share much of their underlying code base, only changing the number and types of radial functions (e.g. adding the treatment of small-component radial functions, which do not exist in the scalar relativistic or two-component spin-orbit coupled cases). For the discussion of the Q4C framework below, a general operator $\hat{F}$ and a general real-space matrix element $F_{ij}$ for the molecule ($F^{uc}_{ij}$ for the periodic solid) are introduced to represent the large or small components of scalar integrals for the Hamiltonian and overlap matrix elements. 

For molecular systems, the corresponding matrix element $F_{ij}$ is evaluated as
\begin{equation}
F_{ij}
=\sum_{\mathbf{r}}w(\mathbf{r}){\varphi_i^{X}}^*(\mathbf{r}) \hat{F} \varphi^{X}_j(\mathbf{r})
\quad \text{for } X = L, S \, ,
\end{equation}
where the summation runs over the overlap region between scalar basis functions $\varphi^{X}_i$ and $\varphi^{X}_j$, 
as illustrated in Fig.~\ref{fig:partition_molecule}, and the indices $i$ and $j$ range up to $N^X_{\text{scalar}}$ for large or small component scalar NAO basis functions. For periodic systems, the integration is confined to the volume of a single reference unit cell. The corresponding real-space unit-cell matrix elements $F^{uc}_{ij}$ are calculated as
\begin{equation}
F^{uc}_{ij}
=\sum_{\mathbf{r}}w(\mathbf{r}){\varphi_{i}^{X}}^*(\mathbf{r}) \hat{F} \varphi^{X}_{j}(\mathbf{r}) \, .
\end{equation} 
These real-space scalar integrals are evaluated by summing up the contributions from grid points located in the overlap areas where the spatial extents of the scalar basis functions $\varphi_i^X$ and $\varphi_j^X$ overlap within the 0-th unit cell, as highlighted by the cyan regions in Fig.~\ref{fig:partition_solids}. Here, the indices $i$ and $j$ range up to $N^{X,\text{uc}}_{\text{scalar}}$. corresponding to the number of scalar basis functions that are non-zero within the central (0-th) unit cell. Importantly, this index represents the complete subset of scalar NAO basis functions that are non-zero within the central (0-th) unit cell, including those whose centered atoms are located in neighboring periodic images. 

In real-space integration, evaluating matrix elements across the entire global integration grid simultaneously is computationally inefficient and scales poorly with system size. Efficient linear scaling integration is achieved by a partitioning scheme illustrated in Fig \ref{fig:partition_molecule} for a molecular system and Fig. \ref{fig:partition_solids} for a periodic solid system.  In this scheme, the global integration points P are first distributed across multiple Message Passing Interface (MPI) tasks. As visualized by the color-coded regions in Fig. \ref{fig:partition_molecule} and Fig. \ref{fig:partition_solids}, the global grid is segmented into sub-regions, with each MPI task exclusively owning a specific portion of the real-space grid. Subsequently, each MPI task further subdivides its assigned subset of points into non-overlapping, local batches of points $B_v$ using a grid-adapted cut-plane algorithm~\cite{HAVU20098367}, where $v$ denotes the index of batches. For CPU execution in FHI-aims, these batches default to 100 points per batch on average, while for GPU calculations (not yet supported by the present Q4C code) the target average batch size is set to 200 points. The global integration for the matrix elements $F_{ij}$ of the molecular system can therefore be rewritten as
\begin{equation}
\begin{split}
F_{ij} &= \sum_{\mathrm{task}} F^{\mathrm{task}}_{ij} \\
&= \sum_{\mathrm{task}} \sum_{B_v \in \mathrm{task}} F_{ij}[B_v] \\
&= \sum_{\mathrm{task}} \sum_{B_v \in \mathrm{task}} \sum_{\mathbf{r} \in B_v} w(\mathbf{r})\, \varphi_i^{X*}(\mathbf{r})\, \hat{F}\, \varphi^{X}_j(\mathbf{r}) \, .
\end{split}
\end{equation}
Extended to periodic boundary conditions, the integration for the $F^{uc}_{ij}$ is expressed as:
\begin{equation}
\begin{split}
F^{uc}_{ij} &= \sum_{\mathrm{task}} F^{uc,\mathrm{task}}_{ij} \\
&= \sum_{\mathrm{task}} \sum_{B_v \in \mathrm{task}} F^{uc}_{ij}[B_v] \\
&= \sum_{\mathrm{task}} \sum_{B_v \in \mathrm{task}} \sum_{\mathbf{r} \in B_v} w(\mathbf{r})\, \varphi_{i}^{X*}(\mathbf{r})\, \hat{F}\, \varphi^{X}_{j}(\mathbf{r}) \, .
\end{split}
\end{equation}
These terms, $F_{ij}[B_v]$and $F^{uc}_{ij}[B_v]$, represent the partial integral contributions over different batches $B_v$ associated with the large or small component scalar basis functions for the matrix elements $F^{\mathrm{task}}_{ij}$ and $F^{uc,\mathrm{task}}_{ij}$ on that task. Because these batches are non-overlapping and uniquely assigned to one specific MPI task, all real-space operations are carried out independently, allowing parallelization to proceed straightforwardly and efficiently. The indices $i$ and $j$ are restricted to run over a reduced-dimensional subspace, $\mathrm{nnz}(B_v)$~\cite{HUHN2020107314, HAVU20098367}, which is defined to contain only the large or small component scalar basis functions that have non-zero values on those specific grid points for any given batch.

\begin{figure}
    \centering
    \includegraphics[width=1.0\linewidth]{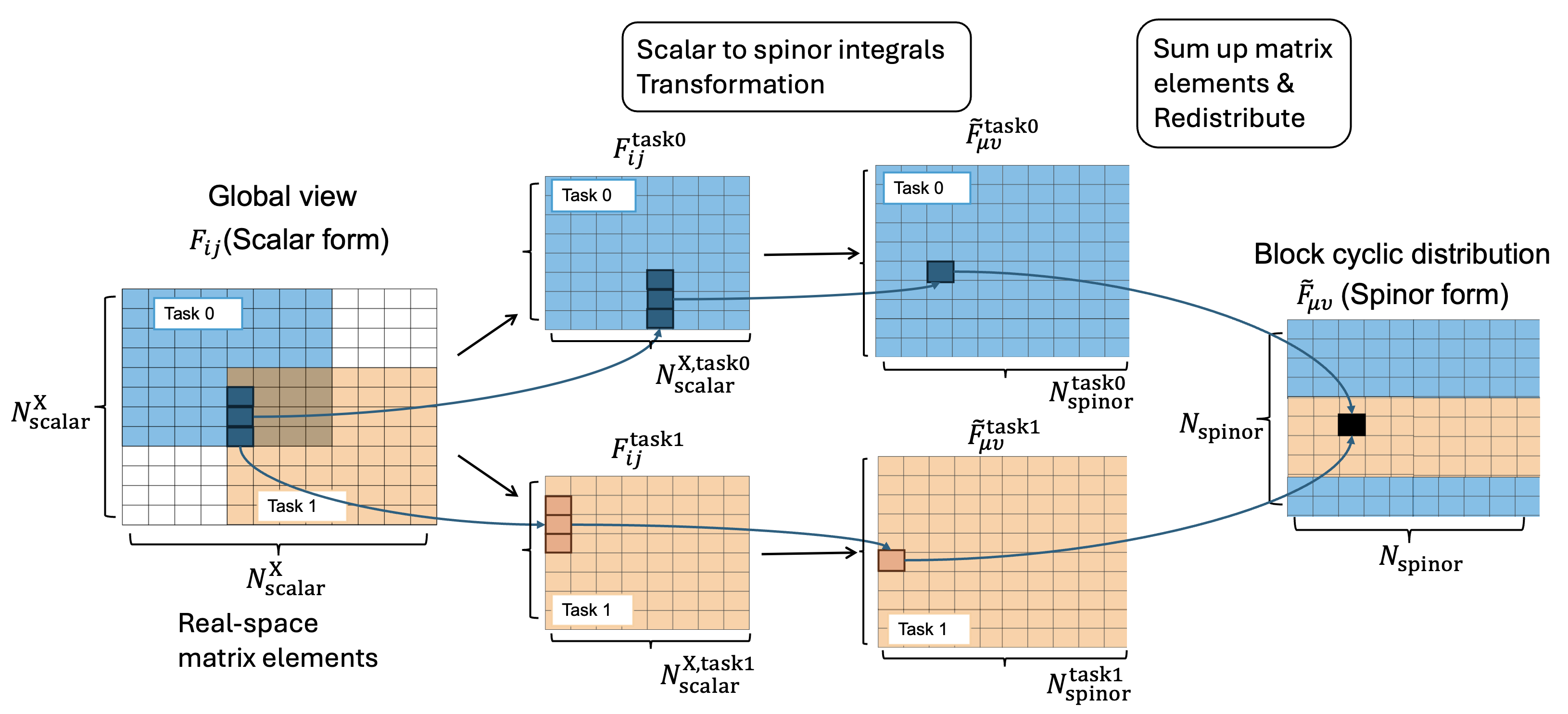}
    \caption{Modified ``locally indexed'' dense storage and transformation scheme for a non-periodic molecule system with two MPI tasks. Each task independently integrates its assigned real-space batches ($B_v$) to construct the locally-indexed dense scalar matrix elements ($F^{\mathrm{task}}_{ij}$). These scalar elements are then transformed into the spinor basis ($\tilde{F}^{\mathrm{task}}_{\mu\nu}$). Finally, the locally-indexed spinor matrix elements are summed across tasks and redistributed into a block-cyclic layout for the ELPA eigensolver. Parts of the integrals that contribute to an exemplary final matrix element (black rectangle, right) are highlighted across different steps of the algorithm.}
    \label{fig:RSDD_molecule}
\end{figure}
\begin{figure}
    \centering
    \includegraphics[width=1.0\linewidth]{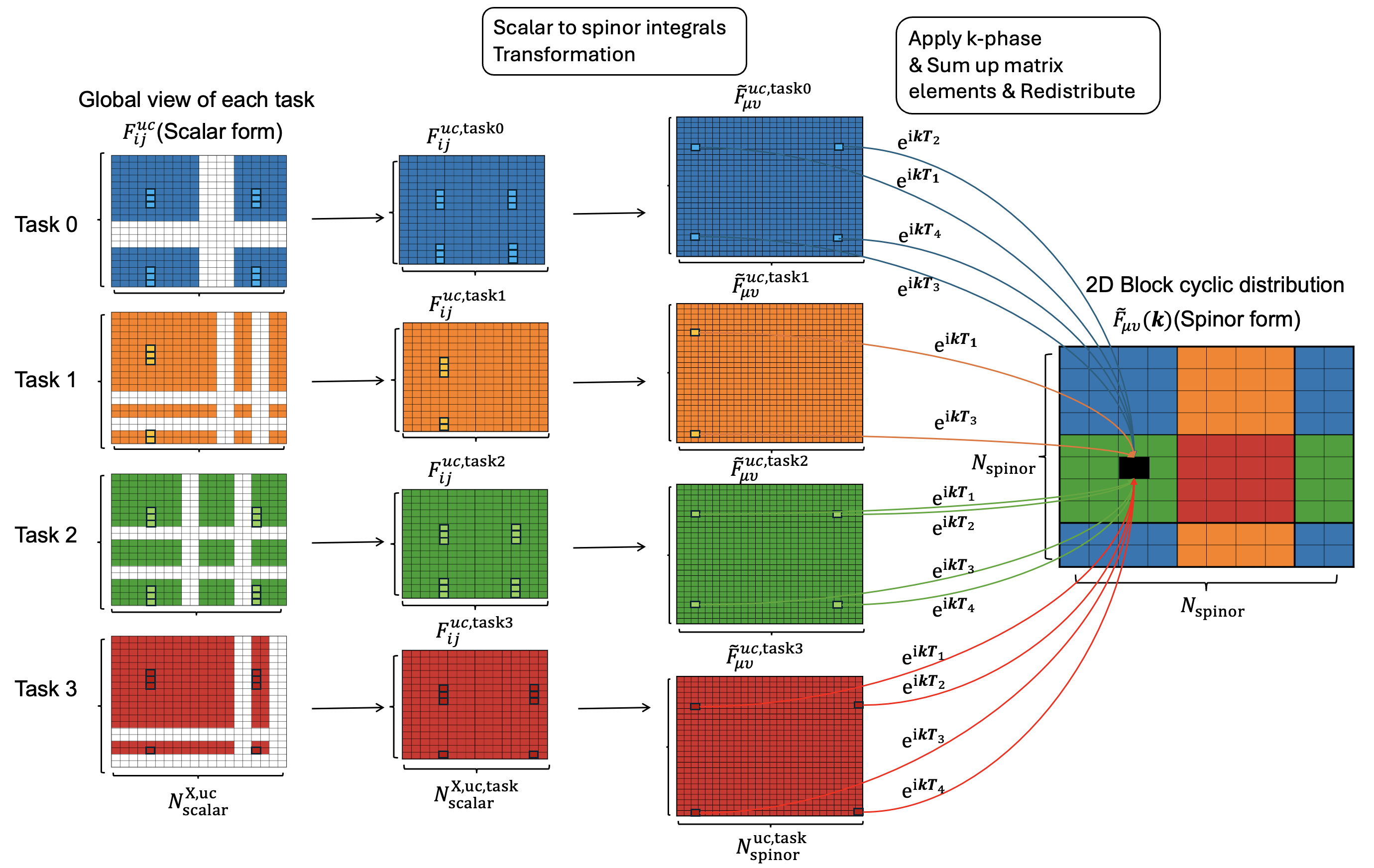}
    \caption{Modified ``locally indexed'' dense storage and transformation scheme for the periodic system with four MPI tasks. Tasks integrate local batches ($B_v$) to build locally-indexed scalar matrices for the reference unit cell ($F^{uc,\mathrm{task}}_{ij}$), which are subsequently transformed to the spinor basis ($\tilde{F}^{uc,\mathrm{task}}_{\mu\nu}$). In the final communication step, the algorithm simultaneously executes the Bloch $\mathbf{k}$-phase summation across unit cells and redistributes the aggregated global spinor matrix elements into a 2D block-cyclic layout. Parts of the integrals that contribute to an exemplary final matrix element (black rectangle, right) are highlighted across different steps of the algorithm.}
    \label{fig:RSDD_solid}
\end{figure}

After these scalar integrals $F^{\text{task}}_{ij}$ and
$F^{uc,\text{task}}_{ij}$ are computed on the individual MPI tasks, they are stored in a distributed format and subsequently transformed into the spinor basis representation of the Q4C formalism. The optimized locally indexed real-space domain decomposition algorithm achieves this process through the modified ``locally indexed'' dense storage and transformation scheme as shown in Fig. \ref{fig:RSDD_molecule} for the molecular case with two MPI tasks and Fig. \ref{fig:RSDD_solid} for the periodic case with four MPI tasks. As noted above, the term ``locally indexed'' refers to the storage, on each MPI task, of only those integration contributions to $F_{ij}$ and $F^{uc}_{ij}$ that are non-zero on the subset of grid batches assigned to that task. These contributions are accumulated in a dense local matrix, which corresponds to a sub-block of the full real-space matrix restricted to the scalar basis functions with non-zero support on the batches handled by that task, as illustrated in the global view of the matrix in the scalar basis form in Fig. \ref{fig:RSDD_molecule} and Fig. \ref{fig:RSDD_solid}.  Different MPI tasks may hold parts of the same overall matrix element, but this does not affect the overall compactness or computational efficiency of the scheme. The memory per task thus remains bounded as system size increases, provided the number of MPI ranks scales accordingly. 

Utilizing this locally-indexed dense matrix, each MPI task performs the transformation of these matrix elements from the scalar form to the spinor form as shown in Fig. \ref{fig:RSDD_molecule} and Fig. \ref{fig:RSDD_solid}. For the molecular system, this gives
\begin{equation}
\begin{split}
\tilde{F}^{\mathrm{task}}_{\mu\nu}  &=\sum_{i,j}^{N^{X,\mathrm{task}}_{\mathrm{scalar}}} ({G^{X}_{i\mu}})^{*}F^{\mathrm{task}}_{ij}G^{X}_{j\nu} \\
&= \sum_{B_v \in \mathrm{task}} \sum_{\mathbf{r} \in B_v}
w(\mathbf{r})\, \left[ \sum_{i}^{N^{X,\mathrm{task}}_{\mathrm{scalar}}} G^{X}_{i\mu}\,\varphi^X_i(\mathbf{r}) \right]^{*} \hat{F} \left[ \sum_{j}^{N^{X,\mathrm{task}}_{\mathrm{scalar}}} G^{X}_{j\nu}\,\varphi^X_j(\mathbf{r}) \right]
\end{split}
\end{equation}
and, for the periodic system,
\begin{equation}
\begin{split}
\tilde{F}^{uc,\mathrm{task}}_{\mu\nu} &=\sum_{i,j}^{N^{X,\mathrm{uc,task}}_{\mathrm{scalar}}} ({G^{X}_{i\mu}})^{*}F^{uc,\mathrm{task}}_{ij}G^{X}_{j\nu} \\
&= \sum_{B_v \in \mathrm{task}} \sum_{\mathbf{r} \in B_v}
w(\mathbf{r})\, \left[ \sum_{i}^{N^{X,\mathrm{uc,task}}_{\mathrm{scalar}}} G^{X}_{i\mu}\,\varphi^X_{i}(\mathbf{r}) \right]^{*} \hat{F} \left[ \sum_{j}^{N^{X,\mathrm{uc,task}}_{\mathrm{scalar}}} G^{X}_{j\nu}\,\varphi^X_{j}(\mathbf{r}) \right] \, .
\end{split}
\end{equation}
These expressions transform each real-valued scalar matrix element into a complex-valued spinor matrix element for the Q4C formalism. Here, $\mu$ and
$\nu$ index the locally-indexed spinor basis functions, and the transformation matrices $G^X_{i\mu}$ contain the Clebsch--Gordan coefficients that combine the scalar basis functions into the spinor basis functions as described in Eq. 14(a)(b).

The procedure to build the locally-indexed spinor basis functions and to construct the corresponding transformation matrices $G^X$ is outlined in Algorithm\ref{alg:1}. In this context, ``globally indexed'' refers to the identical numbering of the basis functions across the entire system for each MPI task while $N^{\mathrm{task}}_{\mathrm{spinor}}$ or ($N^{\mathrm{uc,task}}_{\mathrm{spinor}}$ for periodic systems) denotes the number of locally-indexed spinor basis functions, and the local-to-global spinor index array stores the correspondence between the locally and the globally indexed spinor basis functions. In short, Algorithm~\ref{alg:1} determines which locally-indexed spinor basis functions are present on each MPI task and assembles the transformation matrices $G^X$ directly in the local indexing.

\begin{figure}
\begin{algorithm}[H]
\caption{Construction of the locally-indexed spinor basis and transformation matrices.}
\label{alg:1}
\begin{algorithmic}[1]
\State Build a global relation table that maps each globally-indexed spinor basis function $\chi^X_\mu$ ($\mu = 1, \dots, N_{\mathrm{spinor}}$) to its associated globally-indexed scalar basis functions $\varphi^X_i$ ($i = 1, \dots, N^X_{\mathrm{scalar}}$) of the same component ($X \in \{L, S\}$); for periodic systems, only scalar basis functions with non-zero support in the reference (0-th) unit cell ($i = 1, \dots, N^{X,\mathrm{uc}}_{\mathrm{scalar}}$) are included.
\For{each globally-indexed spinor basis function $\chi^X_\mu$}
    \For{each locally-indexed scalar basis function on the current task ($i = 1, \dots, N^{X,\mathrm{task}}_{\mathrm{scalar}}$ for molecules, or $N^{X,\mathrm{uc,task}}_{\mathrm{scalar}}$ for periodic systems)}
        \If{the global index associated with local index $i$ appears in the relation table entry of $\chi^X_\mu$}
            \State Increment the local spinor basis count $N^{\mathrm{task}}_{\mathrm{spinor}}/N^{\mathrm{uc,task}}_{\mathrm{spinor}}$
            \State Append the global index $\mu$ to the local-to-global spinor index array
        \EndIf
    \EndFor
\EndFor
\State Build a local relation table mapping each locally-indexed spinor basis function to the local indices of its associated scalar basis functions. Assemble the transformation matrices $G^{X,\mathrm{task}}_{i\mu}$ used in the scalar-to-spinor transformations of $\tilde{F}^{\mathrm{task}}_{\mu\nu}$ (molecular) and $\tilde{F}^{uc,\mathrm{task}}_{\mu\nu}$ (periodic).
\end{algorithmic}
\end{algorithm}
\end{figure}

After the scalar-to-spinor integrals transformation, the locally-indexed spinor matrix elements $\tilde{F}^{\mathrm{task}}_{\mu\nu}$ and $\tilde{F}^{\mathrm{uc,task}}_{\mu\nu}$ are distributed across MPI tasks. To prepare these matrices for the ELPA\cite{marek_elpa_2014} eigensolver, they need to be converted into block-cyclic distribution across the processor grid, as shown in Fig. \ref{fig:RSDD_molecule} for a two MPI tasks molecular case and Fig. \ref{fig:RSDD_solid} for a four MPI tasks periodic case. This conversion involves determining the process-to-element ownership mapping and performing point-to-point MPI communication to aggregate contributions from overlapping regions. For molecular systems, the locally-indexed spinor matrix elements from different tasks are summed and simultaneously redistributed into the block-cyclic layout:
\begin{equation}
\tilde{F}_{\mu\nu}=\sum_{\text{task}}\!\bigl(\tilde{F}^{\text{task}}_{\mu\nu}) \, .
\end{equation}
For periodic systems, the aggregation of spinor integral contributions across MPI tasks and the Bloch $\mathbf{k}$-phase summation is executed in the same step, which gives
\begin{equation}
\tilde{F}_{\mu\nu}(\mathbf{k}) = \sum_{\mathrm{task}} {\sum_{\mu^\prime\nu^\prime}}^{'} \exp\left[i\mathbf{k}\cdot (\mathbf{T}_{M(\nu^\prime)}-\mathbf{T}_{M)(\mu^\prime)})\right] \tilde{F}^{uc,\mathrm{task}}_{\mu^\prime\nu^\prime} \, .
\end{equation}
Here, the same restricted sum convention and summation indices are used as described alongside Eq. (\ref{restrictedsum}). Crucially, at no point is the full matrix ever gathered on a single MPI task, ensuring that memory and computational load remain distributed across all processes. Once the block-cyclic distribution and MPI communication are established, the Hamiltonian and overlap matrices in the spinor basis form are distributed across different MPI tasks. Finally, these matrices are passed to the ELPA eigensolver for the generalized eigenvalue problems.

For the density update, the integration sequence proceeds in reverse. Following the solution of the generalized eigenproblems, the density matrix in the spinor basis form is constructed within the block-cyclic distribution across individual MPI tasks. For the molecular case, the density matrix for each task is evaluated as:
\begin{equation}
D^{\mathrm{task}}_{\mu\nu}=\sum_l^{N_{\text{occ}}} f_l ({C^{+,\mathrm{task}}_{\mu l}})^{*} C^{+,\mathrm{task}}_{\nu l}
\end{equation}
where $f_l$	denotes the occupation number. For periodic systems, the occupied eigenvectors at each $k$-point are accumulated with the respective Bloch $k$-phase on each task
\begin{equation}
    D^{\mathrm{task}}_{\mu\nu} =\sum_{\mathbf{k}}\exp[{i\mathbf{k}\cdot(\mathbf{T}_{M(\nu)}-\mathbf{T}_{M(\mu)})}] \sum_l^{N_{\text{occ}}} f_l(\mathbf{k}) [{C^{+,\mathrm{task}}_{\mu l}(\mathbf{k})}]^{*} C^{+,\mathrm{task}}_{\nu l}(\mathbf{k}) \, .
\end{equation}
Using the transformation matrices, the spinor density matrix $D^{\mathrm{task}}_{ij}$ is mapped back to the scalar density matrices $D^{\prime,\mathrm{task}}_{ij}$ for both the large and small components, preserving the block-cyclic distributed format. The resulting scalar density matrix is subsequently redistributed via MPI communication and converted into a locally-indexed dense format. Finally, the real-space electron density is updated from this locally indexed scalar density matrix from each task within each self-consistent field (SCF) iteration. For the molecular system, the density is given by:
\begin{equation}
\begin{aligned}
n(\mathbf r)
&= \operatorname{Re}\sum_{\mathrm{task}}\biggl[\,\sum_{\mu,\nu}^{N^{\mathrm{task}}_{\mathrm{spinor}}} D^{\mathrm{task}}_{\mu\nu} {\chi^{X}_{\mu}}^{*}\chi^{X}_{\nu}\biggr] \\
&=
\operatorname{Re}\sum_{task}\biggl[\,
\sum^{N^{X,\mathrm{task}}_{\mathrm{scalar}}}_{i,j}\Bigl(\sum^{N^{\mathrm{task}}_{\mathrm{spinor}}}_{\mu,\nu} D^{\mathrm{task}}_{\mu\nu}\,{G^{X}_{i\mu}}^{*}G^{X}_{j\nu}\Bigr){\varphi^{X}_{i}}^{*}\,\varphi^{X}_{j}\biggr] \\
&= \operatorname{Re}\sum_{\mathrm{task}}\biggl[\,\sum^{N^{X,\mathrm{task}}_{\mathrm{scalar}}}_{i,j} D^{\prime,\mathrm{task}}_{ij}\,{\varphi^{X}_{i}}^{*}\,\varphi^{X}_{j}\biggr] \, .
\end{aligned}
\end{equation}

Analogously, for periodic systems, the real-space electron density is accumulated by summing the localized contributions from every MPI task. To optimize the evaluation, an intermediate scalar density matrix is explicitly constructed on each task:
\begin{equation}
\begin{aligned}
n(\mathbf r)
&= \operatorname{Re}\sum_{\mathrm{task}}\biggl[\,\sum_{\mu,\nu}^{N^{\mathrm{task}}_{\mathrm{spinor}}} D^{\mathrm{task}}_{\mu\nu} {\chi^{X}_{\mu}}^{*}\chi^{X}_{\nu}\biggr] \\
&=
\operatorname{Re}\sum_{task}\biggl[\,
\sum^{N^{X,\mathrm{uc,task}}_{\mathrm{scalar}}}_{i,j}\Bigl(\sum^{N^{\mathrm{task}}_{\mathrm{spinor}}}_{\mu,\nu} D^{\mathrm{task}}_{\mu\nu}\,{G^{X}_{i\mu}}^{*}G^{X}_{j\nu}\Bigr)
   {\varphi^{X}_{i}}^{*}\,\varphi^{X}_{j}\biggr] \\
&= \operatorname{Re}\sum_{\mathrm{task}}\biggl[\,\sum^{N^{X,\mathrm{uc,task}}_{\mathrm{scalar}}}_{i,j} D^{\prime,task}_{ij}\,{\varphi^{X}_{i}}^{*}\,\varphi^{X}_{j}\biggr] \, .
\end{aligned}
\end{equation}

\subsection{Workflow of the algorithm}

\begin{figure}
    \centering
    \includegraphics[width=1.0\linewidth]{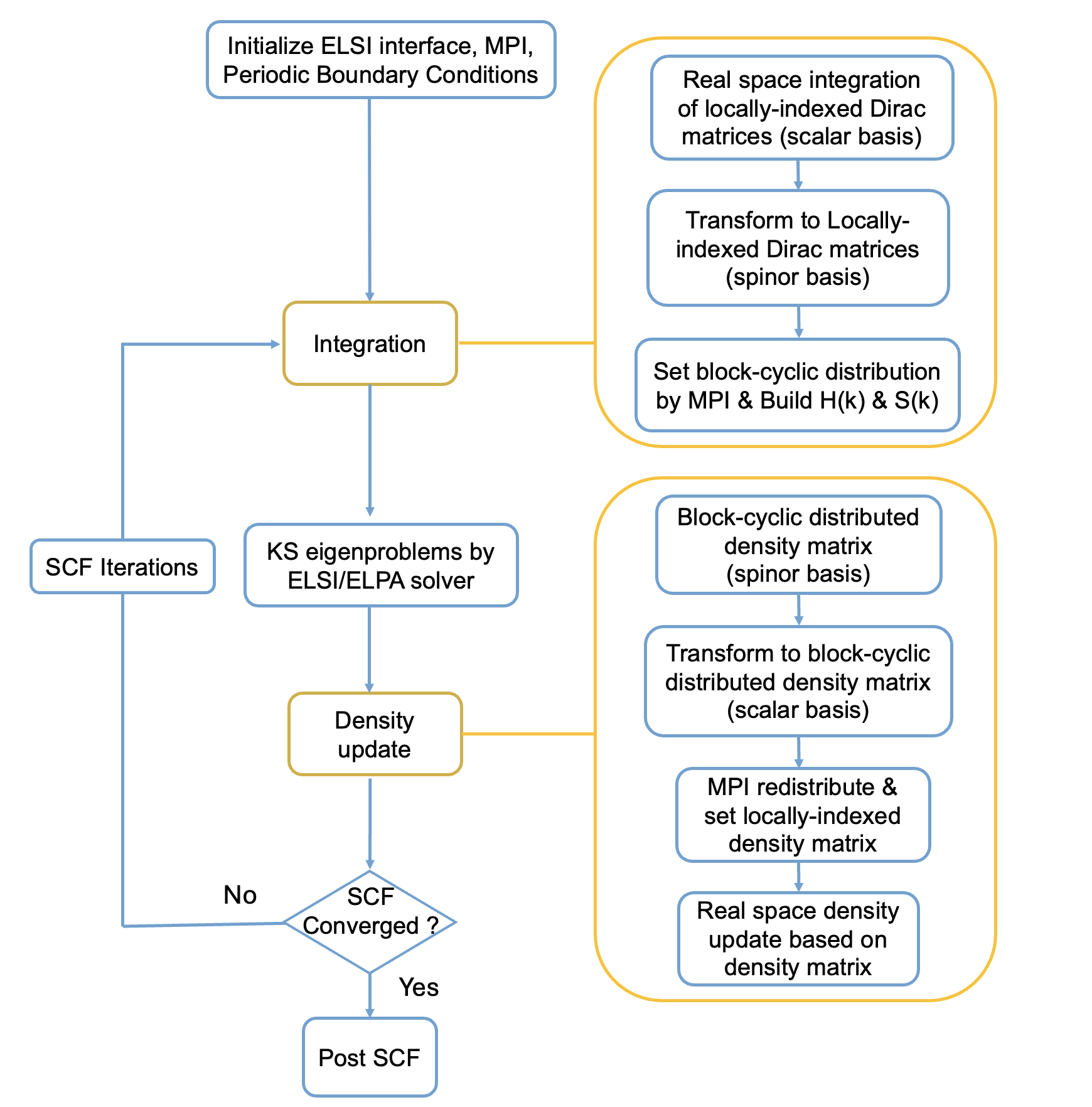}
    \caption{High-level workflow of the large-scale parallel distributed Q4C implementation with periodic boundary conditions in FHI-aims. Blue represents the computational steps. Gold denotes the integration and density-update parts with sub-steps.}
    \label{fig:workflow}
\end{figure}
The high-level workflow for the large-scale Q4C algorithm is shown in Fig~\ref{fig:workflow}. The procedure begins with the initialization setup of the Electronic Structure Infrastructure (ELSI)~\cite{YU2018267, YU2020107459} interface, the Message Passing Interface (MPI), and the periodic boundary conditions for solid systems. The ELSI and MPI initializations are adapted to the Q4C method since the dimensions of the Hamiltonian and overlap matrices differ from those of the scalar-relativistic (SR) and non-relativistic (NR) cases. Regarding periodic boundary conditions, both the large and small component scalar NAO basis functions are constructed together with the corresponding transformation matrices that contain the Clebsch-Gordan coefficients.

For the integration, the locally-indexed real-space domain decomposition algorithm described in the section III A is employed. The "locally-indexed" Dirac Hamiltonian and overlap matrix elements in the scalar basis form are evaluated with the real-space partitioned integration method, then transformed to the spinor basis form, and finally set into a block-cyclic distribution, from which the $k$-dependent Hamiltonian $H(k)$ and overlap $S(k)$ matrices are built.
The generalized KS eigenproblems are subsequently solved with the ELPA solver in the ELSI infrastructure. As illustrated in the density update module of Fig.~\ref{fig:workflow}, the procedure systematically unpacks the eigensolutions generated by the ELPA solver. The sequence initiates with the formulation of the block-cyclic distributed density matrix in the spinor basis.  Using the transformation matrices, the spinor density matrix is mapped back to the scalar density matrices, with the block-cyclic distributed layout preserved. The resulting scalar density matrix is then redistributed via MPI and transformed to the locally indexed dense format. The electron density is finally updated from the locally-indexed scalar density matrix. The iteration repeats until the SCF cycle converges, after which the post-SCF analysis is performed.

\section{Benchmarking Results}

\subsection{Benchmark results for selected periodic solids}

All DFT calculations are performed using the all-electron FHI-aims\cite{BLUM20092175,abbott2026roadmap} code with numeric atom-centered orbital basis sets. These simulations employ the Perdew-Berke-Ernzerhof (PBE) functional \cite{PhysRevLett.77.3865} and use the "intermediate" settings for both the real-space integration grids and the basis sets. We conducted the benchmark on a now-retired infiniband connected partition of the Duke Computer Cluster (DCC), a generalized high-performance computing (HPC) resource at Duke University. Each compute node had 42 cores and contained approximately 467 GB of total physical memory.

\begin{figure}
    \centering
    \includegraphics[width=1.0\linewidth]{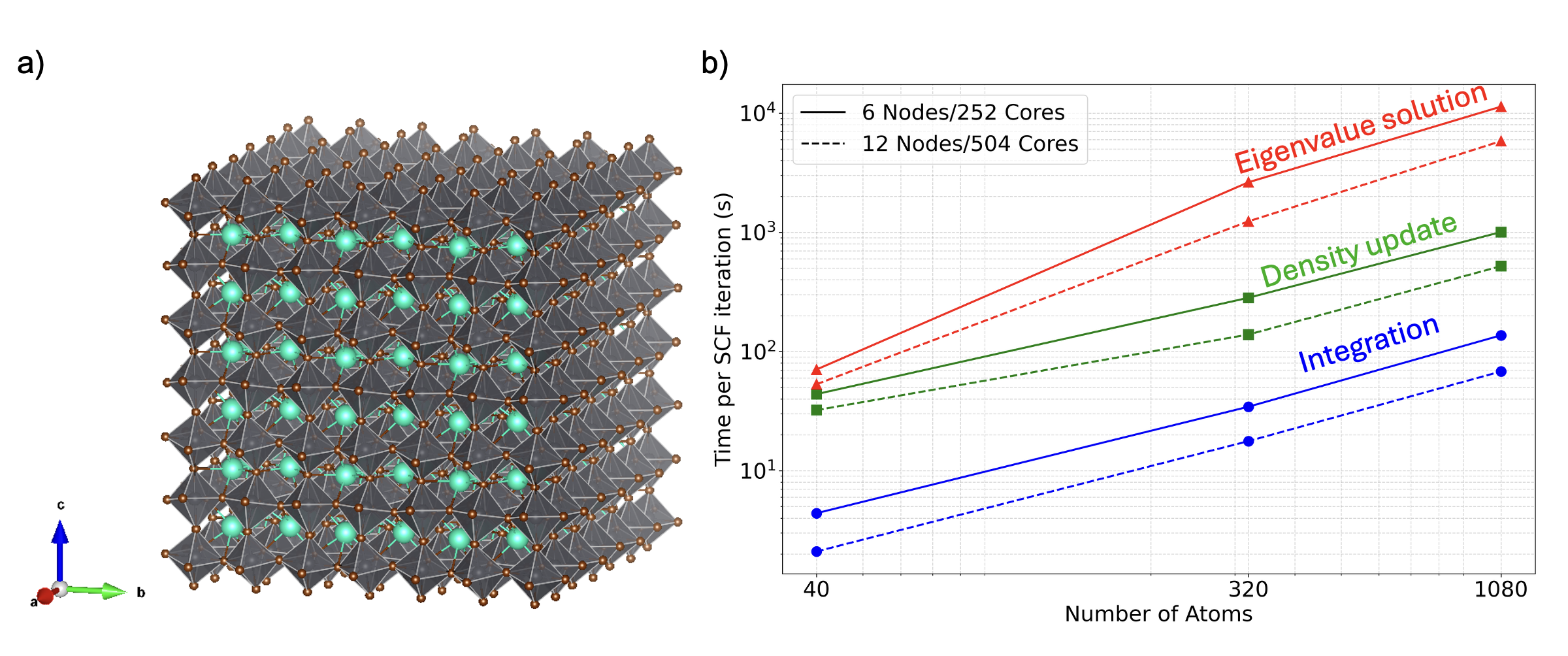}
    \caption{(a) Structure of a 1080-atom CsPbBr$_{3}$ supercell. (b) Time scaling of the large-scale Q4C method per self-consistent field (SCF) iteration for CsPbBr$_{3}$ supercells of 40, 320, and 1080 atoms. Wall times are reported per SCF iteration, separated into the three main components: integration (blue, circles), density update (green, squares), and eigenvalue solution (red,
triangles). Solid lines correspond to runs on 6 nodes (252 MPI tasks) and dashed lines to runs on 12 nodes (504 MPI tasks) of the DCC cluster.}
    \label{fig:scaling}
\end{figure}

To test the scaling of the large-scale Q4C implementation, we select different supercell sizes of the heavy-element containing metal halide perovskite semiconductor  CsPbBr$_{3}$. The detailed unit cell structures are provided in the Supplemental Information. The atomic structure of the largest CsPbBr$_{3}$ supercell with 1080 atoms is shown in Fig~\ref{fig:scaling}a.
We set the k-grid to (1$\times$1$\times$1) for the 1080-atom supercell, k-grid (2$\times$2$\times$2) for the 320-atom supercell, and k-grid (4$\times$4$\times$4) for the 40-atom supercell. We ran the test on 6 nodes (252 MPI tasks) and 12 nodes (504 MPI tasks) of the DCC cluster, chosen to assess scaling within the 16 nodes-per-job limit imposed by the cluster.

Fig~\ref{fig:scaling}b shows the time scaling of the large-scale Q4C method for different operations within one self-consistent field (SCF) iteration with increasing CsPbBr$_3$ supercell sizes. These operations include real-space integration, density update, and solution of the generalized eigenvalue problem. The real-space operations integration and density update reach approximate $\mathcal{O}(N)$ linear scaling. In contrast, the solution to the eigenvalue problem with the ELPA eigensolver exhibits a much steeper scaling in our test cases. Doubling the node count from 6 to 12 yields a near $2\times$ speedup across all three operations for the 320-atom and 1080-atom supercells. However, the 40-atom system attains a reduced speedup of $1.4$--$1.8\times$, as MPI communication becomes comparable to the per-node computational work at this small system size.

\begin{table}
\caption{Benchmark results for the large layered hybrid organic-inorganic perovskite systems with the scalar-relativistic (SR) and quasi-four-component
(Q4C) methods. The reported time represents the average time per Self-Consistent Field (SCF) iteration across various computational operations. Nodes refer to the number of nodes, with each node containing 42 MPI tasks. The number of basis functions for the Q4C method represents the total number of large and small component scalar NAO basis functions.}
\label{tab:performance_comparison}
\begin{ruledtabular}
\begin{tabular}{lcccccc}
\multirow{2}{*}{\textbf{System}} & \multirow{2}{*}{\textbf{Method}} & \multirow{2}{*}{\textbf{Nodes}} & \multirow{2}{*}{\textbf{No.\ of Basis}} & \multicolumn{3}{c}{\textbf{Time [s]}} \\
\cline{5-7}
 & & & & \textbf{Integration} & \textbf{Eigenproblem} & \textbf{Density} \\
\hline
\multirow{4}{*}{\shortstack[c] {\\$\text{C}_{512}\text{H}_{768}\text{Bi}_{1}\text{I}_{128}$\\$\text{N}_{64}\text{Pb}_{31}$(1504 atoms)}}
  & SR  &  8 &  24,096 &  25.7 &   47.4 &  39.7 \\
  & SR  & 16 &  24,096 &  13.2 &   29.9 &  20.7 \\
  & Q4C &  8  &  96,384 &  83.6 &  996.0 & 293.8 \\
  & Q4C & 16 &  96,384 &  43.2 &  535.5 & 149.2 \\
\hline
\multirow{4}{*}{\shortstack[c]{\\$\text{C}_{1152}\text{H}_{1728}\text{Bi}_{2}\text{I}_{288}$\\$\text{N}_{144}\text{Pb}_{69}$(3383 atoms)}}
  & SR  & 8 &  54,157 &  80.1 &  434.2 &  123.5 \\
  & SR  & 16 &  54,157 &  38.8 &  245.3 &  63.0 \\
  & Q4C & 8 & 216,628 & 226.9 & 10231.7 & 1011.0 \\
  & Q4C & 16 & 216,628 & 115.0 & 5318.6 & 516.2 \\
\end{tabular}
\end{ruledtabular}
\end{table}

\begin{table}
\caption{Memory analysis for the largest periodic structures with the large-scale quasi-four-component (Q4C) implementation. Reported are the maximum (Max.) tracked peak memory usage per MPI task and the size of the largest single array allocation, taken across all MPI tasks. Nodes refer to the number of nodes, with each node containing 42 MPI tasks. $D^S$ represents the small component density matrix in the Q4C method. }
\label{tab:memory_analysis}
\begin{ruledtabular}
\begin{tabular}{lcccc}
\multirow{2}{*}{\textbf{System}} & \multirow{2}{*}{\textbf{Nodes}} & \multirow{2}{*}{\textbf{Max. Peak Memory \ [MB]}}  & \multicolumn{2}{c}{\textbf{Largest Array Allocation}} \\
\cline{4-5}
 & & & \textbf{Name} & \textbf{Size [MB]} \\
\hline
\multirow{2}{*}{\shortstack[c]{\\$\text{C}_{512}\text{H}_{768}\text{Bi}_{1}\text{I}_{128}$\\$\text{N}_{64}\text{Pb}_{31}$(1504 atoms)}}
&  8 & 2688.2 & $D^S$ & 559.2 \\
 & 16 & 1693.1 & $D^S$ & 358.9 \\
\hline
\multirow{2}{*}{\shortstack[c]{\\$\text{C}_{1152}\text{H}_{1728}\text{Bi}_{2}\text{I}_{288}$\\$\text{N}_{144}\text{Pb}_{69}$(3383 atoms)}}
 &  8 & 5585.2 & $D^S$ & 944.6 \\
 & 16 & 3096.2 & $D^S$ & 537.3 \\
\end{tabular}
\end{ruledtabular}
\end{table}
\raggedbottom

Furthermore, to investigate the reach of the implementation for extremely large system sizes from a past production example,\cite{PRXEnergy.2.023010} we perform benchmark calculations on Bi-doped 2D Ruddlesden--Popper hybrid organic--inorganic perovskites based on phenylethylammonium lead iodide (PEPI). We first consider a Bi-doped PEPI supercell containing 1,504 atoms (later called 1504-PEPI for short), in which a single Pb$^{2+}$ site is replaced by a Bi$^{3+}$ dopant. Subsequently, we extend the benchmarks to a larger defect complex PEPI supercell comprising 3,383 atoms (later called 3383-PEPI), in which two Bi$^{3+}$ dopants together with a charge-compensating Pb$^{2+}$ vacancy substitute for three Pb$^{2+}$ ions. These two materials are adopted from Ref.~\cite{PRXEnergy.2.023010}, and the crystal structures of them are shown in Fig.~\ref{fig:band_structures}a and Fig.~\ref{fig:band_structures}e, respectively. For both materials, we use the $1\times1\times1$ k-grid and the "intermediate" numerical settings. We perform the DFT-PBE calculations with the atomic ZORA scalar relativistic (SR) method\cite{BLUM20092175,abbott2026roadmap} and the relativistic Q4C method on 8 (336 MPI tasks) and 16 nodes (672 MPI tasks) of the DCC cluster.

Table \ref{tab:performance_comparison} reports the timing of one self-consistent field (SCF) iteration for the two large hybrid organic--inorganic perovskite systems with SR and Q4C level of relativistic theory. The number of basis functions required for the Q4C method is four times larger than that of the SR method for the same structure. The computational total time per SCF iteration is decomposed into three components: numerical integration, solution of the generalized eigenproblem, and density update.  Consistent with the trend already observed for the scaling of CsPbBr$_3$ supercells, the eigenproblem dominates the time for both systems with the Q4C method, while the real-space operations contribute a smaller fraction. In addition, doubling the number of nodes from 8 to 16 reduces the time by nearly a factor of 2, demonstrating the strong scaling behavior of the Q4C implementation. Comparing the SR and Q4C methods for the identical structures and the same number of nodes in Table \ref{tab:performance_comparison}, the average time cost increases at different rates across all three operations. The real-space integration time increases by $2.8$--$3.3\times$ for both systems, scaling sublinearly with the number of basis functions. Meanwhile, the density update time is $7.2$--$8.2\times$ longer, due to the construction of both large and small components of the density matrices in the Q4C method. In contrast, the time for the generalized eigenproblem grows by a factor of $17.9$--$23.6\times$ across all four cases.

Moreover, memory analysis for the large-scale Q4C implementation reveals the scalability and parallel efficiency. The detailed results for the peak memory usage per MPI task and the largest array allocation are summarized in Table~\ref{tab:memory_analysis}. The small component density matrix $D^S$ is the dominant largest array allocation per MPI task, accounting for roughly 15--20\% of the peak per-task memory. Decreasing the node counts from 16 to 8 effectively reduces the maximum peak memory by nearly 55--63\% across all MPI tasks for these systems, which indicates the parallel capabilities of the distributed memory architecture. We note that, while these systems reflect the limit of what was tested in this work, the overall results do not yet indicate that the scaling limit of the implementation has been exceeded in terms of system size and of MPI rank count. Thus, it is conceivable that even larger systems may be enabled by the implementation in the future, should they be required.

\subsection{Electronic band structure for hybrid organic--inorganic perovskites}

\begin{figure}
    \centering
    \includegraphics[width=1.0\linewidth]{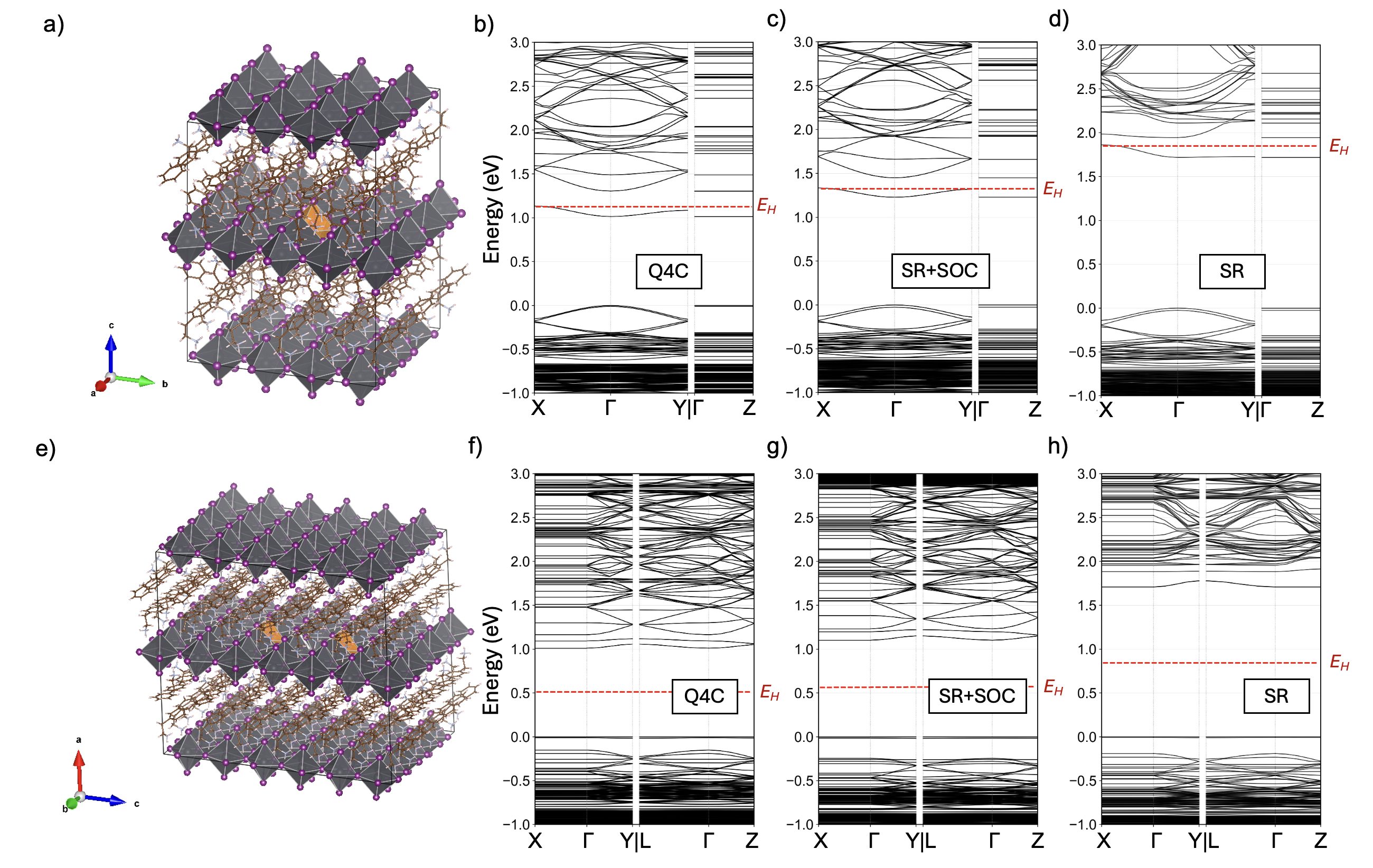}
    \caption{%
        Crystal structures and electronic band structures of two PEPI
        supercells computed with the PBE functional at three levels of relativistic theory.
        Top row [(a)--(d)]: 1{,}504-atom supercell
        Bottom row [(e)--(h)]: 3{,}383-atom supercell.
        (a) Structure of 1{,}504-atom supercell
        (e) Structure of 3{,}383-atom supercell
        (b),(f) Scalar relativistic (SR) method (atomic ZORA).
        (c),(g) SR with non-self-consistent second-variational spin--orbit
        coupling (SR+SOC).
        (d),(h) Self-consistent quasi-four-component (Q4C) treatment.
        Dashed red horizontal lines labeled $E_H$ indicate the chemical potential, above which states are effectively unoccupied.%
    }
    \label{fig:band_structures}
\end{figure}

One direct application of the scalable Q4C implementation is the investigation of the electronic band structures of complex defect supercell systems. In hybrid organic--inorganic perovskites containing heavy elements, relativistic effects play an important role as spin-orbit coupling (SOC) strongly shapes the energy bands. The electronic band structures of the 1504-PEPI and 3383-PEPI supercells are calculated using the PBE functional across three relativistic methods: atomic ZORA\cite{BLUM20092175,abbott2026roadmap} scalar relativistic (SR), SR with non-self-consistent SOC,\cite{PhysRevMaterials.1.033803} and the Q4C method, which incorporates self-consistent SOC, all relevant basis functions that reflect the atomic behavior near the nucleus, and makes no shape approximation to the SOC operator. The k-space grid is set to (2$\times$2$\times$2) for the 1504-PEPI structure and (1$\times$1$\times$1) for the 3383-PEPI structure. The "intermediate" settings for the basis functions and integration grids are utilized. The crystal structures are shown in Figs. ~\ref{fig:band_structures} (a)(e).

Fig ~\ref{fig:band_structures} indicates the critical role of relativistic effects in the electronic band structures.  For the 1504-PEPI supercell, unlike the SR band structure from Fig. \ref{fig:band_structures} (d), the second-order variational (non-self-consistent) SOC Fig.\ref{fig:band_structures} (c) shows that the inclusion of SOC reduces the band gap, shifts the conduction bands to Pb-derived bands, which are the upper bands above $E_H$, and introduces band splitting for the conduction band above the $E_H$. The quasi-four-component (Q4C) from Fig.\ref{fig:band_structures} (b) (with self-consistent SOC) shows a further decrease in the band gap. This modification particularly affects the position and character of the Bi-derived defect state, which is the band below the $E_H$. 

For the 3383-PEPI supercell, the overall trends established for the 1504-PEPI supercell carry over: SOC reduces the gap and induces splittings near the band edges (Fig.~\ref{fig:band_structures}(g)). However, the Q4C treatment (Fig.~\ref{fig:band_structures}(h)) further refines the position of the energy levels in the valence band relative to the SR+SOC result. This refinement highlights a fundamental limitation of the two-step, non-selfconsistent SR+SOC approach.\cite{PhysRevMaterials.1.033803} In scalar relativistic treatments, the basis functions lack the fine-structure splitting inherent to fully relativistic quantum mechanics; they do not possess separate radial functions for the $j=l+s$ and $j=l-s$ states within a given orbital angular momentum $l$ channel. Consequently, applying non-self-consistent SOC on top of SR basis functions implies that the basis set is not fully converged for SOC effects---for instance, it lacks the distinct $p_{1/2}$ radial functions that are crucial for accurately describing heavy element Bi, which dominates the upper valence bands. This effect is well known, e.g., in linearized augmented plane wave calculations and corrected for $p$ orbitals in the Wien2k code.\cite{10.1063/1.5143061} Consistent with our own past benchmark work,\cite{PhysRevB.103.245144} the Q4C method naturally generates and includes these separate spin-orbit coupled radial functions from the outset, yielding a more complete basis and a physically accurate description of the defect states and the valence band structure.

To the best of our knowledge, the 3383-PEPI supercell is among the largest fully self-consistent four-component relativistic DFT band-structure calculations reported to date, and it confirms that the Q4C implementation presented here remains numerically stable and parallel-efficient at the few-thousand-atom scale required for realistic defect studies in heavy-element semiconductors.

\section{Conclusions}

In this work, an efficient, parallel implementation of the quasi-four-component (Q4C) relativistic DFT with numeric atom-centered orbitals for molecular and solid systems is presented. The implementation was performed in the FHI-aims code, but the algorithms are generally applicable, demonstrating how far one can take Becke's seminal, overlapping atom-centered grid integration scheme.\cite{10.1063/1.454033} The relativistic four-component formulation within the NAO framework is presented comprehensively. We optimized and developed the locally-indexed real-space domain decomposition algorithm for the real-space operations (i.e., integration and density update). Following the modified locally-indexed dense storage and transformation scheme for Q4C integration, the Hamiltonian and overlap matrices are set to a block-cyclic distribution layout, which is then sent to the massively parallel, dense eigensolver ELPA. The implementation enables large-scale calculations with system sizes exceeding 3000 atoms. We perform the benchmarking results for both computational time scaling and distributed memory efficiency, and compare with the SR method.

Several directions are are still needed for the future development of the Q4C method within the NAO framework. The first is the extension to open-shell, spin-polarized systems, which will enable Q4C calculations of magnetic materials and systems with unpaired electrons. The second is the implementation of analytical total energy gradients (forces), which is an essential next step for geometry optimization, ab initio molecular dynamics, and phonon calculations of heavy-element-containing compounds at the fully relativistic level. Another aspect is to extend the Q4C method with the efficient hybrid density functional approach \cite{doi:10.1021/acs.jctc.1c00137,Kokott_2024}, e.g., the HSE06 functional, to combine the accurate treatment of relativistic effects with an improved description of exchange interactions in heavy-element semiconductors.

\section*{Acknowledgments}
V.B. and W.Z. were supported by the U.S. National Science Foundation (NSF) under Award Number 2323803. R.Z.'s work on the scalar to spinor transformation was supported by the NSF under Award Number 1729297, supervised by V.B.

\section*{Data Availability}
All data supporting this work are available in the NOMAD database at \url{10.17172/nomad.f8ky-0e5r}. The deposited data contains input structures for all calculations, real-space integration grid examples for molecular and periodic systems, scaling results for the CsPbBr$_3$ benchmark supercells, and data from the large-scale PEPI calculations, including memory-usage measurements and electronic band structures. Detailed descriptions and direct URLs for the individual datasets in NOMAD are provided in the Supplemental Information.

\section*{Supplementary Material}

The Supplementary Material file includes atomic structure, computational details, and data repository locations of the benzene, graphene, CsPbBr$_3$ and doped phenethylammonium lead iodide based example structures used for illustration and benchmark purposes in the paper.

\appendix
\section{Relation of Clebsch-Gordon coefficients and complex spherical harmonics}
\label{CG}
 The spin-angular functions\cite{dyall2007relativistic} of a single electron wave function can be written as
\begin{equation}
    \Upsilon_{\kappa,m_j}=\sum_{m_l,m_s}\Braket{lm_lsm_s|jm_j}\Tilde{Y}_l^{m_l}(\Omega)\eta(m_s) \, .
\end{equation} 
Here, $\Braket{lm_lsm_s|jm_j}$ is the Clebsch-Gordan coefficient coupling orbital 
angular momentum $l$ and spin $s=\tfrac{1}{2}$ to total angular momentum $j$, $\tilde{Y}_l^{m_l}(\Omega)$ are the complex spherical harmonics, and the two-component Pauli spinor  $\eta(m_s)$ is defined as
\begin{equation}
    \eta\!\left(+\tfrac{1}{2}\right) = \begin{pmatrix}1\\0\end{pmatrix}, \qquad
    \eta\!\left(-\tfrac{1}{2}\right) = \begin{pmatrix}0\\1\end{pmatrix}.
\end{equation}
Since $m_s = \pm\tfrac{1}{2}$ fixes $m_l = m_j - m_s$, the definition of the Clebsch-Gordan coefficient $C_\alpha, C_\beta$ from Eq. (10) is
\begin{equation}
    C_\alpha \equiv
    \Braket{l,\,m_j{-}\tfrac{1}{2};\,\tfrac{1}{2},\,{+}\tfrac{1}{2}\,|\,j,\,m_j},
    \qquad
    C_\beta  \equiv
    \Braket{l,\,m_j{+}\tfrac{1}{2};\,\tfrac{1}{2},\,{-}\tfrac{1}{2}\,|\,j,\,m_j}.
\end{equation}

\section{Real-valued spherical harmonics}
\label{RSH}
With the Condon-Shortley phase convention \cite{BLANCO199719}, the real-valued spherical harmonics can be expressed as

\begin{equation}
Y_{l}^{m} = \begin{cases}
(-1)^m\sqrt{2}K_{l}^{m}cos(m\phi)P_{l}^{m}(cos\theta)  & \text{if } m > 0 \\
K_{l}^{m}P_{l}^{m}(cos\theta) & \text{if } m = 0 \\
(-1)^m \sqrt{2}K_{l}^{m}sin(|m|\phi)P_{l}^{|m|}(cos\theta)  & \text{if } m < 0.
\end{cases}
\end{equation}
Here, $\theta$ and $\phi$ are angular spherical coordinates, $P_{l}^{m}$ are the associated Legendre polynomials

\begin{equation}
    P_{l}^{m}(cos\theta)=(sin\theta)^m\frac{d^m}{d(cos\theta)^m}(P_{l}(cos\theta))
\end{equation}
where $P_{l}$ is a Legendre polynomial, and $K_{l}^{m}$ is the normalization factor, which is 

\begin{equation}
    K_{l}^{m}=\sqrt{\frac{(2l+1)(l-|m|)!}{4\pi(l+|m|)!}} \, .
\end{equation}

\bibliography{references}

\end{document}


\preprint{}

\title{A Large-scale Parallel Implementation of Quasi-Four-Component Relativistic Density Functional Theory with Numeric Atom-centered Orbitals \\ Supplemental Information} 



\author{Wentao Zhang}
\affiliation{Thomas Lord Department of Mechanical Engineering and Materials Science, Duke University, Durham, North Carolina 27708, United States}

\author{Rundong Zhao}
\affiliation{Thomas Lord Department of Mechanical Engineering and Materials Science, Duke University, Durham, North Carolina 27708, United States}

\author{Volker Blum}
\affiliation{Thomas Lord Department of Mechanical Engineering and Materials Science, Duke University, Durham, North Carolina 27708, United States}
\affiliation{Department of Chemistry, Duke University, Durham, North Carolina 27708, United States}


\date{\today}

\pacs{}

\maketitle 


\section{Real-space integration grid}

For the real-space integration grid shown in Fig.~1, we select a single benzene molecule from ref \cite{pubchem_benzene} and convert the structure information to the \texttt{geometry.in} input format that is used in FHI-aims. Then, we perform the DFT calculation with the PBE functional and large-scale Q4C algorithm using the "light" numerical settings with four MPI tasks. The detailed coordinates files for each MPI task can be accessed at \url{https://nomad-lab.eu/prod/v1/gui/user/datasets/dataset/id/cOUE9hu5S-COrZ9SGI3S0Q/entry/id/IQewmT4SX198j6uId6WUj4aUSJ9c}.
The \texttt{geometry.in} file for the benzene molecule is:
\begin{verbatim}
atom -1.213100 -0.688400 0.000000 C
atom -1.202800 0.706400 0.000100 C
atom -0.010300 -1.394800 0.000000 C
atom 0.010400 1.394800 -0.000100 C
atom 1.202800 -0.706300 0.000000 C
atom 1.213100 0.688400 0.000000 C
atom -2.157700 -1.224400 0.000000 H
atom -2.139300 1.256400 0.000100 H
atom -0.018400 -2.480900 -0.000100 H
atom 0.018400 2.480800 0.000000 H
atom 2.139400 -1.256300 0.000100 H
atom 2.157700 1.224500 0.000000 H
\end{verbatim}

To demonstrate the integration grid in Fig.~2 for a periodic system, a graphene sheet is modeled. The initial structure is constructed from a primitive two-atom hexagonal cell utilizing the in-plane lattice constant of a = 2.46 Å (corresponding to a C–C bond length of 1.42 Å from ref\cite{PhysRev.100.544}), and a vacuum spacing of 40 Å is applied along the out-of-plane direction. A supercell transformation is then applied to generate an 8×8 supercell of graphene consisting of 128 atoms. The calculation is performed with DFT-PBE with Q4C using 4 MPI tasks with the "light" numerical settings. The detailed coordinate files for each MPI task and the corresponding global batch index information are available at \url{<https://nomad-lab.eu/prod/v1/gui/user/uploads/upload/id/CzPRiaXdQiOoUF8PDVhAoA>}. The \texttt{geometry.in} file for the 8x8 graphene is generated based on the following unit cell:
\begin{verbatim}
lattice_vector   2.460000   0.000000   0.000000
lattice_vector   1.230000   2.130400   0.000000
lattice_vector   0.000000   0.000000  40.000000

atom   0.000000  0.000000   0.000000  C
atom   1.230000   0.710133   0.000000  C
\end{verbatim}

\section{Benchmark Details}
\subsection{Time scaling for the CsPbBr$_{3}$ supercells}

We choose the 20-atom cell of CsPbBr$_{3}$ from Ref.~\cite{doi:10.1021/cg400645t}. as the
starting structure. A supercell transformation is applied to convert the
tetragonal cell into a cubic 40-atom cell. The 320-atom ($2\times2\times2$) and
1080-atom ($3\times3\times3$) supercells are then generated from this
40-atom cell. All data underlying the timing and scaling analysis are
available at \url{<https://nomad-lab.eu/prod/v1/gui/user/uploads/upload/id/YRFAOJxRSyGlB-jy9LkOfA>}.
The geometry specifications of all structures are provided in the FHI-aims \texttt{geometry.in} format. The geometry file for the 40-atom unit cell of CsPbBr$_{3}$ is given below.
\begin{verbatim}
lattice_vector 8.244000 0.000000 8.198200
lattice_vector 0.000000 11.735100 0.000000
lattice_vector -8.244000 0.000000 8.198200

atom 0.000000 5.867550 0.000000 Pb
atom 4.122000 5.867550 4.099100 Pb
atom 0.000000 0.000000 0.000000 Pb
atom 4.122000 0.000000 4.099100 Pb
atom -3.866436 2.933775 8.255751 Cs
atom -0.255564 8.801325 12.354851 Cs
atom 3.866436 8.801325 8.140649 Cs
atom 0.255564 2.933775 4.041549 Cs
atom -5.824633 5.587198 9.887603 Br
atom 1.702633 6.147902 5.788503 Br
atom 5.824633 11.454748 6.508797 Br
atom -1.702633 0.280352 10.607897 Br
atom 5.824633 6.147902 6.508797 Br
atom -1.702633 5.587198 10.607897 Br
atom -5.824633 0.280352 9.887603 Br
atom 1.702633 11.454748 5.788503 Br
atom 0.030503 8.801325 0.373018 Br
atom 4.091497 2.933775 4.472118 Br
atom -0.030503 2.933775 16.023382 Br
atom -4.091497 8.801325 11.924282 Br
atom 0.000000 5.867550 8.198200 Pb
atom -4.122000 5.867550 4.099100 Pb
atom 0.000000 0.000000 8.198200 Pb
atom -4.122000 0.000000 4.099100 Pb
atom 4.377564 2.933775 8.255751 Cs
atom -0.255564 8.801325 4.156651 Cs
atom -4.377564 8.801325 8.140649 Cs
atom 0.255564 2.933775 12.239749 Cs
atom 2.419367 5.587198 9.887603 Br
atom 1.702633 6.147902 13.986703 Br
atom -2.419367 11.454748 6.508797 Br
atom -1.702633 0.280352 2.409697 Br
atom -2.419367 6.147902 6.508797 Br
atom -1.702633 5.587198 2.409697 Br
atom 2.419367 0.280352 9.887603 Br
atom 1.702633 11.454748 13.986703 Br
atom 0.030503 8.801325 8.571218 Br
atom -4.152503 2.933775 4.472118 Br
atom -0.030503 2.933775 7.825182 Br
atom 4.152503 8.801325 11.924282 Br
\end{verbatim}

\subsection{PEPI with defect}

The 1504-PEPI is based on the Bi-doped  4x4 phenylethylammonium lead iodide, and the 3383-atom is based on the (6x6) PEPI with vacancy.
The two PEPI structures are adopted from Ref.~\cite{PRXEnergy.2.023010}. The reference data for the scaling test and the memory usage reported in the table and all data related to the corresponding band structures are available at \url{<https://nomad-lab.eu/prod/v1/gui/user/uploads/upload/id/5PZx2K_2TUuDJ4kOlLWsRw>}.


%